\documentclass[twocolumn, numberedappendix]{openjournal}
\usepackage[T1]{fontenc}
\usepackage{graphicx} 
\usepackage{graphicx}	
\usepackage{amsmath}	
\usepackage{amssymb}	
\usepackage{color}
\usepackage{xcolor}

\usepackage{hyperref}
\hypersetup{
	colorlinks=true,
	urlcolor=blue,    
	linkcolor=black,  
	citecolor=blue,   
}
\usepackage{orcidlink}

\usepackage{bm}
\usepackage{enumerate}

\newcommand{\LC}{\mathrm{LC}}
\newcommand{\R}{\mathcal{R}}

\begin{document}
	\title{Repeating partial disruptions and two-body relaxation \vspace{-4em}}

	\author{Luca Broggi \orcidlink{0000-0002-9076-1094}$^{1,2}$}
	\email[E-mail: ]{luca.broggi@unimib.it}
	\author{Nicholas C. Stone \orcidlink{0000-0002-4337-9458}$^3$}
        \author{Taeho Ryu \orcidlink{0000-0003-2012-5217}$^{4,5}$}
	\author{Elisa Bortolas \orcidlink{0000-0001-9458-821X}$^{1,2}$}
	\author{Massimo Dotti \orcidlink{0000-0002-1683-5198}$^{1,2}$}
	\author{Matteo Bonetti \orcidlink{0000-0001-7889-6810
}$^{1,2}$}
	\author{Alberto Sesana \orcidlink{0000-0003-4961-1606}$^{1,2}$}

    \affiliation{$^1$Dipartimento di Fisica G. Occhialini,
		Università degli Studi di Milano-Bicocca, piazza della Scienza 3, Milano, Italy.}
	\affiliation{$^2$ INFN, Sezione di Milano-Bicocca, Piazza della Scienza 3, I-20126 Milano, Italy}
	\affiliation{$^3$Racah Institute of Physics, The Hebrew University, 91904, Jerusalem, Israel}
    \affiliation{$^4$Max Planck Institute for Astrophysics, Karl-Schwarzschild-Str.~1, 85748 Garching, Germany}
    \affiliation{$^5$ Physics and Astronomy Department, Johns Hopkins University, Baltimore, MD 21218, USA}

\begin{abstract}
    Two-body relaxation may drive stars onto near-radial orbits around a massive black hole, resulting in a tidal disruption event (TDE). In some circumstances, stars are unlikely to undergo a single terminal disruption, but rather to have a sequence of many grazing encounters with the black hole.  It has long been unclear what is the physical outcome of this sequence: each of these encounters can only liberate a small amount of stellar mass, but may significantly alter the orbit of the star. We study the phenomenon of repeating partial tidal disruptions (pTDEs) by building a semi-analytical model that accounts for mass loss and tidal excitation. In the empty loss cone regime, where two-body relaxation is weak, we estimate the number of consecutive partial disruptions that a star can undergo, on average, before being significantly affected by two-body encounters. We find that in this empty loss cone regime, a star will be destroyed in a sequence of weak pTDEs, possibly explaining the tension between the low observed TDE rate and its higher theoretical estimates.
\end{abstract}

\section{Introduction}
A massive black hole (MBH) can reveal itself and increase its mass via a so-called stellar tidal disruption event (TDE, \citealt{hillsPossiblePowerSource1975,Rees1988, Phinney1989, Evans1989}). A TDE occurs when a star gets too close to the MBH so that the MBH tidal shear overcomes the star self-gravity, resulting in the disruption and accretion of the star by the MBH \citep[e.g.][]{Rees1988, Lodato2009}. Such events produce extremely bright electromagnetic flare observable up to cosmological distances \citep{bloomPossibleRelativisticJetted2011, cenkoSwiftJ205805162012,leloudasSuperluminousTransientASASSN15lh2016}. In particular, they can probe  the dynamical properties of the galactic nuclei in which they are generated \citep[see e.g. the recent reviews by][]{Stone2020,WeversRyu2023}. The most ubiquitous and likely dominant mechanism able to generate a TDE is stellar two-body relaxation: repeated two-body interactions among stars slowly affect the orbital properties of an unlucky star that as a consequence may get too close to the MBH and find itself destroyed. Theoretical works focusing on this aspect typically assess the occurrence rate of TDEs given the properties of the MBH and host environment \citep[e.g.][]{SyerUlmer1999,Magorrian1998,WangMerritt2004}; the growing body of TDE observations \citep[now of the order of $\sim 100$,][] {gezariTidalDisruptionEvents2021} has sparked growing interest in assessing  TDE rates and has recently allowed comparisons between theoretically predicted TDE rates and observationally measured ones. In particular, recent comparisons between theoretical \citep{stoneRatesStellarTidal2016,Stone2020} and observational  \citep{vanVelzen18,Sazonov+2021,Lin22,Yao23} TDE rate estimates suggest that current detection rates are a factor of a few to an order of magnitude lower than empirically-calibrated predictions from classical loss cone theory (though see also \citealt{Roth+21, Polkas+23, Teboul+24}).  This problem may become worse considering the over-representation of observed TDEs in rare post-starburst host galaxies \citep{frenchTidalDisruptionEvents2016,French2020}, pushing down the intrinsic TDE rate in ``normal'' galaxies.

Theoretical TDE rate calculations usually start from a kinetic theory treatment of the statistical evolution of stellar systems \citep{chandrasekharPrinciplesStellarDynamics1942}.  Idealising the stellar distribution as a statistical continuum, one finds that a conical region in velocity space, the ``loss cone,'' is where TDEs occur, and the rate of TDEs is set by the rate of stellar diffusion into this loss cone \citep{frankEffectsMassiveBlack1976, lightmanDistributionConsumptionRate1977,shapiroStarClustersContaining1978, cohnStellarDistributionBlack1978}. This theory relies on the assumption that the disruption of a star happens as soon as the pericentre of its orbit is smaller than
\begin{equation}
    r_\mathrm{TDE} = R_\star \, \left(\frac{M_\bullet}{m_\star}\right)^{1/3}
\end{equation}
known as the \textit{tidal disruption radius}\footnote{A similar approach is also applied to the formation of extreme mass ratio inspirals \citep[EMRIs;][]{hilsGradualApproachCoalescence1995}, which occur when compact objects are captured by MBHs through gravitational wave emission \citep{hopmanOrbitalStatisticsStellar2005,merrittDynamicsEvolutionGalactic2013,broggiExtremeMassRatio2022}. The EMRI loss cone is not as simple as the classical TDE loss cone, however, because of the slow evolution of EMRI orbits over many pericentre passages; in this way, detailed treatments of EMRI dynamics prefigure the aim of this paper.}. Here $m_\star$ and $R_\star$ respectively represent the stellar mass and radius, while $M_\bullet$ is the MBH mass.

In standard loss cone theory, $r_{\rm TDE}$ is typically assumed to be the threshold separation below which the total disruption of the star  occurs, while stars approaching with larger pericentres are assumed to remain unperturbed. However, reality is more complex: the star can either be fully destroyed at a pericentre closer than $r_{\rm TDE}$, or, if passing by at  larger separations, only some outer layers of the envelope may be "peeled off" the star and accreted by the MBH, while a stellar remnant survives the interaction, giving rise to a partial disruption \citep[e.g.][]{Mainetti2017,Ryu2020c}. Such partial disruptions can repeat if the orbit of the star is not dramatically affected by either the partial disruption process itself or by subsequent two-body scatterings over the course of the next orbit.
A number of recent works have addressed the expected emission and accretion features associated with the partial disruption of a single star approaching the MBH \citep{Guillochon2013,Coughlin2019,Miles2020}, and a growing body of literature is reporting observational candidates for partial disruptions, which manifest themselves as repeating events over timescales of $\sim 0.3-3$ yrs \citep{Payne+2021,Wevers2023,Liu2023,Liu2023b,Malyali+2023,Malyali+2023b,somalwar2023}.
There are a number of hypotheses about the origins of stars on extremely bound orbits, including the tidal break up of stellar binaries \citep{Cufari+22}, the migration of stars through a gaseous disk during the last AGN episode \citep{metzgerInteractingStellarEMRIs2022}, or two-body relaxation in super massive black holes binaries \citep{melchorTidalDisruptionEvents2024}.
However, the theoretical modeling of the event rates associated to partial TDEs is still in its infancy \citep{Stone2020, Krolik+2020,ChenShen2021, Zhong+2022, bortolasPartialStellarTidal2023}.

In a recent work, \citet{bortolasPartialStellarTidal2023} explore for the first time the rate of both partial and total TDEs by extending standard loss cone theory to include partial TDEs, with critical pericentres determined from fully relativistic hydrodynamics simulations involving realistic main sequence stars \citep{Ryu2020a,Ryu2020b,Ryu2020d}. In particular, they show that it is possible to relate the rate of total disruptions with the rate of  (repeated) partial disruptions that are expected to occur in nuclear clusters; partial disruptions were found to exceed the rate of total disruptions by a factor of ten or more, especially if the stellar system was dominated by the  empty loss cone regime (see definition in Sec.~\ref{sec:fokkerplanck} below). In \citet{bortolasPartialStellarTidal2023}, the loss cone radius was assumed to be the one at which partial disruptions start to occur, instead of the commonly assumed full disruption radius $r_{\rm TDE}$. However, this first work did not take into account the fact that when a star undergoes multiple partial disruptions, its specific energy can vary owing to the mass stripping it experiences at pericentre, as well as tidal excitation that can be converted into orbital energy.

In this work we present a simple effective model for repeated partial disruptions accounting for the energy changes the star may experience at each pericentre; we estimate the number of repeated partial disruptions stars can experience and their final fate. Furthermore, we assess whether the classical loss cone theory  remains valid when also considering  partial disruptions and, if this is the case, how it could be adapted to properly capture this phenomenon.

The paper is organised as follows: in Section 2 we present the physical phenomena included in our model of partial disruptions, and we summarise the key hypotheses and results of classical loss cone theory as presented in \citet{cohnStellarDistributionBlack1978}; in Section 3 we present the results of some simulations that show the effect of repeated disruptions on orbits unaffected by relaxation; finally, in Section 4 we summarise our results and discuss their implications. In Appendix A we show the results of our simulations when accounting for additional effects.

\section{Partial disruptions and loss cone theory}

Here we build a simple model of partial disruptions to identify and characterise the key features of the process: the change in stellar mass and specific binding energy. We start by describing how the mass loss experienced by partial disruptions can affect their subsequent orbits.

\subsection{Partial Disruptions}
We consider a star of mass $m_\star$ and radius $R_\star$ undergoing a partial disruption around a MBH of mass $M_\bullet$. The star is originally on an orbit with specific energy $E$ and specific angular momentum $J$; in our notation, the orbit is bound when $E > 0$. Since we expect no significant density of stars inside $r_\mathrm{TDE}$, the local potential is Keplerian and the pericentre of the orbit for $J^2 \ll G^2\, M_\bullet^2/(2 E)$ -- as we expect for a tidal disruption -- is
\begin{equation}
    r_\mathrm{p} = \frac{J^2}{2 \, G M_\bullet} \,.
\end{equation}
When the star approaches the central MBH on an orbit with $r_\mathrm{p} \sim r_\mathrm{TDE}$, it loses part of its mass and the strong tidal forces can make its structure oscillate. These two effects generally alter the orbit of the remnant (if it survives the first pericentre passage) as follows.

The process of partial disruption around an MBH strongly affects the specific orbital energy $E$ of the subject star, while its angular momentum change is significantly smaller. The orbit of the surviving stellar core is modified by the receipt of a core kick produced by asymmetric mass loss in the envelope; this asymmetry creates a ``rocket effect'' altering the orbital elements of the core \citep{manukianTurbovelocityStarsKicks2013}. This is in agreement with what found by \citet{Ryu2020c} through hydrodynamical simulations and is consistent with theoretical expectations \citep[see e.g.][]{ivanovNewModelTidally2001}. At the order of magnitude level, the magnitude of the kick imparted to the surviving core, $\delta v_{\rm ml}$, can be written as a fraction $\epsilon$ of the typical velocity at the pericentre $v_p \simeq \sqrt{2\, G\, M_\bullet / r_\mathrm{p}}$, we therefore have

\begin{equation}
    \frac{\left\lvert\delta E_\mathrm{ml} \right\rvert}{E} \simeq \epsilon\, \frac{4\,a}{r_\mathrm{p}}\qquad
    \frac{\left\lvert\delta J_\mathrm{ml}\right\rvert}{J} \simeq \epsilon
\end{equation}
where $a \simeq G\, M_\bullet \, /\, 2 E$ is the semimajor axis of the orbit. Since we focus on very eccentric orbits, $a \gg r_\mathrm{p}$ and the relative change in angular momentum is negligible with respect to the relative change in energy. In addition to the effect of core kicks, a close pericentre passage induces tidal oscillations at the expenses of orbital kinetic energy and orbital angular momentum. Assuming that the oscillations have specific energy $\left \rvert \delta E_\mathrm{tid}\right\rvert = \kappa_E \, G \, m_\star/R_\star$ and specific angular momentum $\left\lvert\delta J_\mathrm{tid}\right\rvert = \kappa_J \, \sqrt{G\, m_\star \, R_\star}$ (with $\kappa_E\sim\kappa_J \lesssim 1$), it follows that for a star on an orbit with $r_\mathrm{p} \sim r_\mathrm{TDE}$
\begin{equation}
    \frac{\left \rvert \delta E_\mathrm{tid}\right\rvert}{E} \simeq \kappa_E \, \left( \frac{m_\star}{M_\bullet} \right)^{2/3}\; \frac{2\, a}{r_p}\qquad \frac{\left\lvert\delta J_\mathrm{tid}\right\rvert}{J} \simeq \kappa_J \, \left( \frac{m_\star}{M_\bullet} \right)^{2/3}.
\end{equation}
showing that the relative change in orbital angular momentum coming from tidal oscillations may be neglected for the very eccentric orbits we are considering.

The mass lost in the process of a partial disruption depends on the orbital parameters (in particular on the pericentre distance), and on the internal structure of the star. We will consider two reference cases:
\begin{itemize}
 \item \textbf{Model A}: a star with mass $m_A = 1 \, M_\odot$ and radius $R_\star = 1\, R_\odot$, $n=3.0$, $r_\mathrm{TDE} = 3.6\cdot 10^{-6}$ pc.\\
 \item \textbf{Model B}: a star with mass $m_B = 0.3 \, M_\odot$ and radius $0.38\, R_\odot$, $n=1.5$, $r_\mathrm{TDE} = 2.0 \cdot 10^{-6}$ pc.
\end{itemize}
where we report the index $n$ of the polytropic model that better agrees with the equilibrium MESA profile \citep{paxtonMODULESEXPERIMENTSLAR2010} for the two cases.
\begin{figure}
 \includegraphics[width=\linewidth]{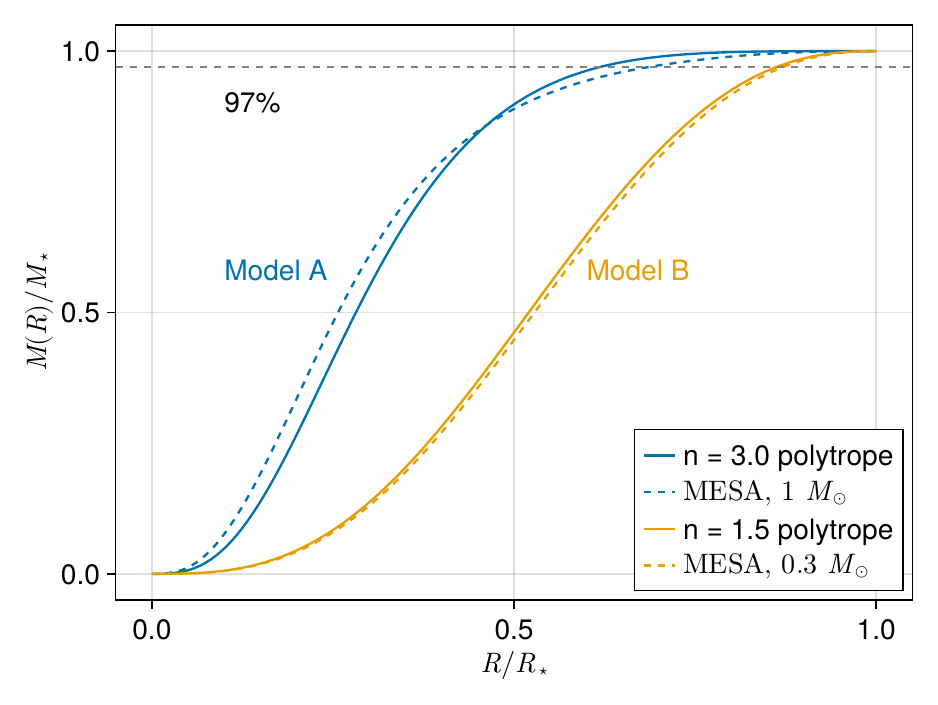}
 \caption{\label{fig:mass_plytropic}Fraction of mass $M(r)$ enclosed in a radius $r$ for model A ($m_\star = 1\, M_\odot$, $R_\star=1\,R_\odot$, blue) and model B ($m_\star = 0.3\, M_\odot$, $R_\star=0.38\,R_\odot$, orange). The distance from the centre is expressed in units of the stellar radius of the star. For each stellar model we show the polytropic model (solid line) and the MESA profile (dashed line) evolved for half its main-sequence lifetime. Model A stars show a more concentrated profile compared to model B stars. Consequently, the same mass loss fraction will produce stronger tidal oscillations for model B (we mark with a dashed line the 3\% mass loss reference). }
\end{figure}

In the case of model A, we estimate the mass loss as a function of the pericentre through the fitting formulae obtained by \citet{Ryu2020a,Ryu2020b,Ryu2020c} for stars on nearly radial orbits ($e \gtrsim 0.99$)
\begin{equation}\label{eq:mass_loss_ryu}
\begin{split}
    \frac{\delta m_A}{m_\star} &= \left(\eta^A_\mathrm{TTDE} \, \frac{r_{TDE}}{r_\mathrm{p}}\right)^\zeta\\
    \log_{10}\zeta &= 0.3 + 3.15\times 10^{-8} \left( \log_{10}\frac{M_\bullet}{1\,M_\odot}\right)^{8.42} \,.
\end{split}
\end{equation}
where $\eta^A_\mathrm{TTDE}$ is a form factor of order unity that determines the pericentre $\eta_\mathrm{TTDE}\, r_\mathrm{TDE}$ corresponding to $\delta m = m_\star$ (i.e. a total disruption).  When considering a model A star:
\begin{equation}\label{eq:eta_ftde}
\begin{split}
    \eta^A_\mathrm{TTDE}(M_\bullet) = \left(0.80+0.26\sqrt{\frac{M_\bullet}{10^6M_\odot}}\right)\times\\
    \left(\frac{1.47 + \exp{\left(2.416\right)}}{1+2.34\, \exp{\left(2.416\right)}}\right).
\end{split}
\end{equation}
In the case we are considering, $M_\bullet = 4\times10^6 M_\odot$ and $\zeta = 3.56$, so that $\eta^A_\mathrm{TTDE}= 0.61$.

In the case of model B, we rely on the simulations by \citet{guillochonHydrodynamicalSimulationsDetermine2013}, that simulated stars with polytropic index $n = 1.5$. Here the mass loss is given by\footnote{The expression of $\delta m/m$ matches the accurate version presented in the erratum (Eq. A9).}
\begin{equation}\label{eq:mass_loss_guillochon}
    \frac{\delta m_B}{m} =  \exp\left(\frac{3.1647 - 6.3777\, \beta + 3.1797\,\beta^2}{1 .- 3.4137\, \beta + 2.4616\, \beta^2}\right)
\end{equation}
for $\beta = r_\mathrm{TDE}/r_\mathrm{p} > 0.5$. We extrapolate this function at larger pericentre values (i.e. smaller $\beta$s) as
\begin{equation}
\begin{split}
    \frac{\delta m_B}{m} = \exp\left( A + B\, \ln \beta + C\, \ln^2\beta \right)\quad \beta < 0.5\\
    A=-68.9227 \quad B=-232.6214 \quad C=-209.6639\\
\end{split}
\end{equation}
that is a smooth extrapolation of the main solution up to the second derivative in the log-log space. Although the simulations of \citet{guillochonHydrodynamicalSimulationsDetermine2013} were Newtonian, this is an acceptable approximation for partial disruptions around lower-mass MBHs such as our case. The value of the pericentre such that $\delta m / m = 1$ corresponds to $\eta^B_\mathrm{TTDE} = 1.11$.

The change in orbital energy due to the ejection of material can be crudely estimated from the asymmetry in the tidal field as \citep[Eq. 32]{stoneConsequencesStrongCompression2013,metzgerInteractingStellarEMRIs2022}
\begin{equation}\label{eq:mass_loss_Stone}
	\delta E_\mathrm{ml}^{S} \simeq - \frac{GM_\bullet}{r_\mathrm{TDE}} \, \left(\frac{R_\star}{r_\mathrm{TDE}}\right)^2 \, \frac{\Delta m}{m - \Delta m} \, .
\end{equation}
The aforementioned equation assumes a pericentre close to the tidal disruption radius $r_\mathrm{TDE}$. A more accurate extension to larger pericentres that agrees with hydrodynamical simulations of a star undergoing a partial disruption around an intermediate MBH is simply obtained by replacing $r_\mathrm{TDE}$ with the pericentre of the orbit \citep{kremerHydrodynamicsCollisionsClose2022, kirogluPartialTidalDisruptions2023}
\begin{equation}\label{eq:mass_loss_Kremer}
	\delta E_\mathrm{ml}^K \simeq - \frac{GM_\bullet}{r_\mathrm{p}} \, \left(\frac{R_\star}{r_\mathrm{p}}\right)^2 \, \frac{\Delta m}{m - \Delta m} \, .
\end{equation}
An alternative estimate, obtained not from physical reasoning but rather from fitting rational functions to simulations of a star being destroyed by an MBH has been obtained by \citet{gaftonRelativisticEffectsTidal2015}
\begin{equation}\label{eq:mass_loss_gafton}
    \delta E_\mathrm{ml}^G \simeq -\frac{G\, m_\star}{R_\star} \left[1 - \left(1-\frac{\delta m}{m_\star}\right)^2\right]^{1.765}\left(1-\frac{\delta m}{m}\right)^{-0.393}.
\end{equation}

\begin{figure*}
    \centering
    \includegraphics[width=0.49\linewidth]{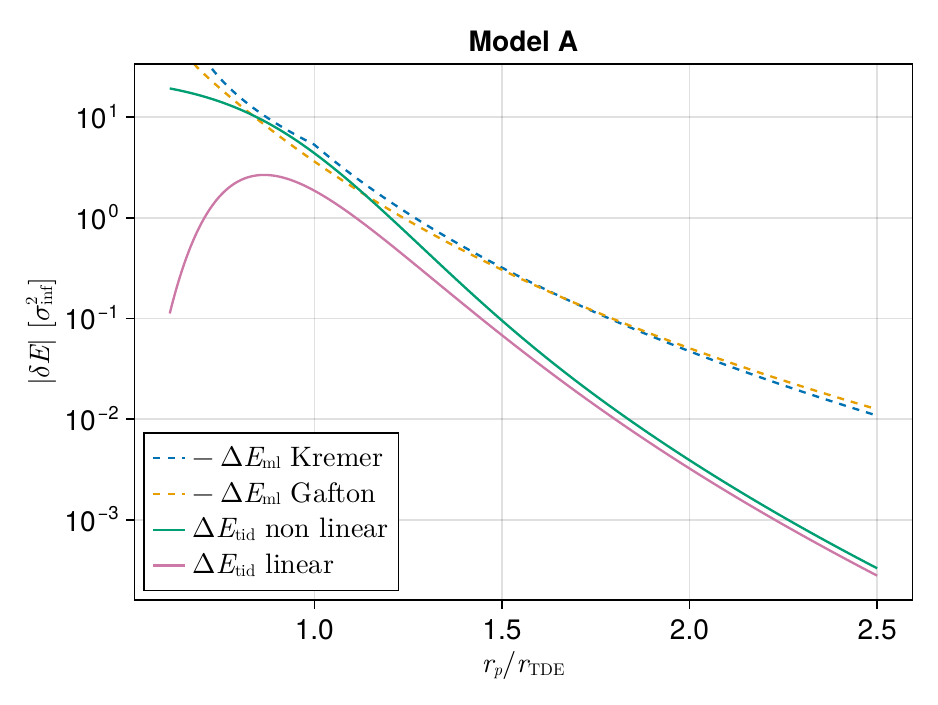}\includegraphics[width=0.49\linewidth]{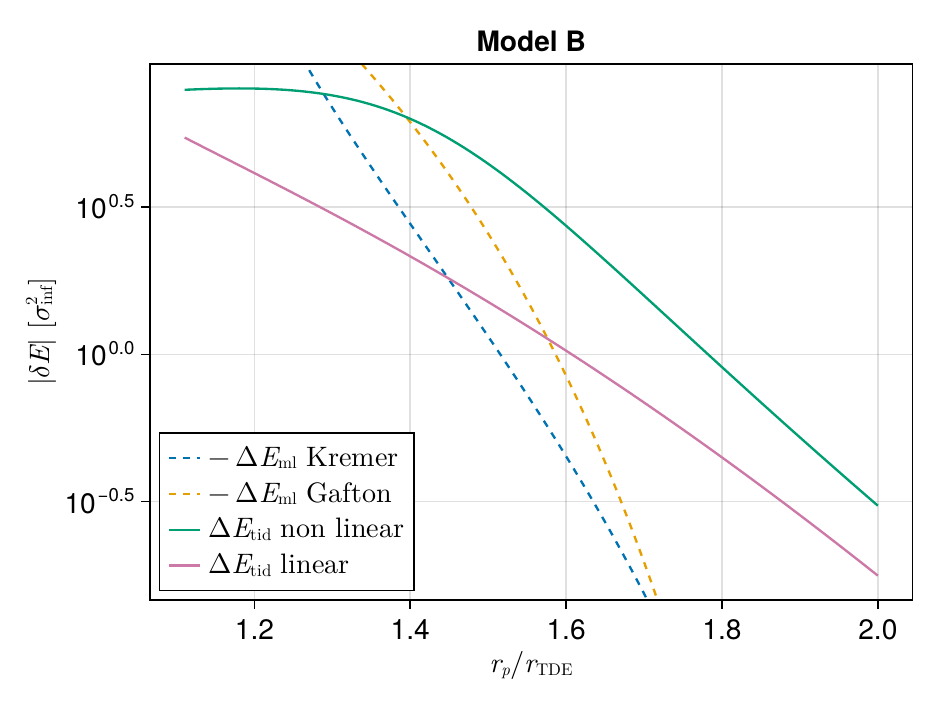}\\
    \caption{Magnitudes of energy changes because of asymmetric mass loss (the ``rocket effect'') and tidal excitation of oscillation modes for the two stellar models we consider, as a function of the normalized pericentre distance. For each model, we compute the mass lost at a given pericentre according to Eq.~\ref{eq:mass_loss_ryu} (model A) or Eq.~\ref{eq:mass_loss_guillochon} (model B) and estimate the corresponding energy change according to Eq.~\ref{eq:mass_loss_Kremer} (dashed blue) or Eq.~\ref{eq:mass_loss_gafton} (dashed orange). For a direct comparison, we also plot the magnitude of tidal oscillations according to the non linear theory Eq.~\ref{eq:nonlin_tid} (solid green) and the linear expansion (solid pink) Eq.~\ref{eq:lin_tid}. Asymptotically, at large $r_p$, the energy change for model A stars is dominated by mass loss, while for model B it is dominated by tidal excitation.}
    \label{fig:competing_effects}
\end{figure*}

The two prescriptions differ more when $\delta m$ is large, while they are comparable (within a factor of 3) when $\delta m \to 0$; in Fig.~\ref{fig:competing_effects} they can be compared directly.

As we already mentioned, tidal forces may also alter the stellar structure, inducing oscillations of the star by converting part of its orbital kinetic energy into the mechanical energy of stellar normal modes. This process, known as tidal excitation  (or the ``dynamical tide''), increases the orbital binding energy in competition with the effects of mass loss. The specific binding energy converted into oscillations can be written as \citep{pressFormationCloseBinaries1977}
\begin{equation}\label{eq:nonlin_tid}
    \delta E_\mathrm{tid} = \frac{G\, M_\bullet^2}{m_\star \, R_\star}\, \left( \frac{R_\star}{r_\mathrm{p}} \right)^6 \, T(\beta)
\end{equation}
where $T(\beta)$ is a dimensionless function that depends on the details of the stellar structure. If the star is already oscillating, whether tidal deformation injects or subtracts energy to the stellar orbit depends on the phase of the ongoing oscillation \citep{laiDynamicalTidesRotating1997}, but Eq.~\ref{eq:nonlin_tid} still gives the approximate amplitude of the perturbation.
Assuming that the oscillations are weak, one can treat them in the linear regime, and the function $T(\beta)$ can be expanded, giving
\begin{equation}\label{eq:lin_tid}
    \delta E_\mathrm{tid}^\mathrm{lin} = \frac{G\, M_\bullet^2}{m_\star \, R_\star} \left( \frac{R_\star}{r_\mathrm{p}} \right)^6 \left[ T_2(\beta) + \left( \frac{R_\star}{r_\mathrm{p}} \right)^2 \!\!T_3(\beta) + \dots\right]
\end{equation}
where $T_2$, $T_3$ are the lowest order terms (quadrupolar and octupolar, respectively) in the linear expansion of the star's response to the tidal field \citep{leeCrossSectionsTidal1986}. Although these functions involve complicated integrals over the internal structure of the star, we employ the analytic fitting functions of \citet[Eq. 7 and Tab.1, with $\eta \simeq \beta^{-3/2}$ since $m_\star \ll M_\bullet$]{portegieszwartQuickMethodCalculating1993}, which are accurate for the polytropic models we use.

When mass loss becomes significant and oscillations are strong, the linear regime becomes inadequate; one must instead solve directly the hydrodynamics of the oscillations. While this is most accurately done with hydrodynamical simulations \citep{manukianTurbovelocityStarsKicks2013}, the tidal coupling driving nonlinear oscillations can be approximated reasonably well with the ``extended affine model'' of \citet{ivanovNewModelTidally2001}, a generalisation of the classical affine model for tidally disrupting stars \citep{luminetDynamicsAffineStar1986}\footnote{The affine model approximates the interior of a star subject to strong tidal perturbations as a set of concentric \citep[but not necessarily co-aligned;][]{ivanovNewModelTidally2001} ellipsoidal shells with time-evolving axis ratios.  In this sense, it can be understood as a nonlinear theory of the quadrupolar tide, neglecting higher-order terms.}. In the case of some polytrope models, analytic fits of the extended affine model's predictions for nonlinear $T(\beta)$ have been presented in \citet[Eq. B1 and B2, $\eta \simeq \beta^{-3/2}$]{generozovOverabundanceBlackHole2018}.
In Fig.~\ref{fig:competing_effects}
we show the competing effects of mass loss and tidal excitation as a function of the pericentre distance for both models. Since stars described by model A are more centrally concentrated with respect to stars of model B -- see Fig.~\ref{fig:mass_plytropic} -- 
tidal excitation is of greater relative importance in model B ($n=1.5$). 

The asymptotic trend at large $r_p$ is different fro the two models: model A stars always decrease their orbital binding energy, because of the dominant effect of asymmetric mass loss.  In contrast, model B stars can increase or decrease their orbital binding energy, because of the dominant effect of tidal excitation.
The energy changes presented here allow us to infer whether a star can undergo multiple partial disruptions, and whether it becomes more or less bound at each passage.

\newpage

\subsection{Two-body relaxation and loss cone theory}\label{sec:fokkerplanck}

The dynamical evolution and relaxation of a nuclear cluster due to two-body encounters is simply and accurately described by the orbit-averaged Fokker-Planck equation \citep{cohnStellarDistributionBlack1978}. By directly computing the expected stochastic variations of a test star's orbital parameters, it can be shown that relaxation in squared angular momentum (normalised to the value $J^2_c$ at the circular orbit of energy $E$) $\R \equiv J^2/J_c^2$ is more efficient than relaxation in $E$ for eccentric orbits. One can identify two time scales \citep{merrittDynamicsEvolutionGalactic2013}:
\begin{itemize}
	\item $T_\mathrm{r}(E) \propto \sigma^3(a)/\rho(a) $ is the typical relaxation time for the energy of a particle with semi major axis $a$, where $\sigma(a)$ and $\rho(a)$ are the velocity dispersion and the density of the stellar distribution at a distance $a$ from the centre, respectively.\\
	\item $\Delta t_\mathrm{rlx}\simeq \R\,T_\mathrm{r}$ is the typical relaxation time for the angular momentum of the particle.
\end{itemize}
Since $0<\R<1$, nearly radial orbits (where stars about to be destroyed are typically located) relax much faster in angular momentum, effectively decoupling the relaxation processes in $\R$ and $E$ for small $\R$. This fact has been exploited by \citet{cohnStellarDistributionBlack1978} to derive a suitable boundary layer solution for the orbit-averaged Fokker-Planck equation in $(E, \R)$ and to consistently include loss cone effects in the equation. Moreover, they derived the expected relaxed distribution of $\R$ among all the particles with energy $E$, including those that will end up in a tidal disruption. We will briefly describe this derivation, in order to highlight the underlying assumptions.

Since $r_\mathrm{p}$ is small compared to the scale radius of the stellar distribution, the value of $\mathcal{R}_{\rm TDE}$ (the dimensionless angular momentum corresponding to $r_{\rm p}=r_{\rm TDE}$) is very small for most of the stars in the distribution. Therefore, one typically assumes that the energy is constant and $E$ can be treated as a parameter. The orbit-averaged Fokker-Planck equation for the phase space distribution function $f(E,\R)$ reduces to
\begin{equation}\label{eq:reduced_oaFP}
     \frac{\partial}{\partial t} f(E,\R, t) = \frac{\partial}{\partial \R} \left[ \R \, D \, \frac{\partial}{\partial \R} f(E,\R, t) \right]
\end{equation}
where $D=D(E)$ is the orbit averaged diffusion coefficient, that for very eccentric orbits characterises the effects of 2-body encounters \citep{merrittDynamicsEvolutionGalactic2013}
\begin{equation}
    P \, D = \langle \Delta \R \rangle_\mathrm{orb} \simeq \frac{1}{2}\frac{\langle \Delta \R^2 \rangle_\mathrm{orb}}{\R} \, .
\end{equation}
Here $P(E)$ is the orbital period (we neglect its weak dependence on $\R$ coming from the stellar potential), while $\langle \Delta \R \rangle_\mathrm{orb}$ and $\langle \Delta \R^2 \rangle_\mathrm{orb}$ are the orbit-averaged expectation values of perturbations in $\mathcal{R}$ and $\mathcal{R}^2$, which depend negligibly on the properties of the star being scattered.\footnote{The advective terms are proportional to the mass of the body being scattered, but they are increasingly negligible as $\R$ goes to zero.} It is useful to quantify the strength of angular momentum diffusion by defining:
\begin{equation} \label{eq:q_def}
    q(E) = \left. \frac{\langle \Delta \R \rangle_\mathrm{orb}}{\R}\right \rvert_{\R_\mathrm{TDE}} = \frac{P \, D}{\R_\mathrm{TDE}} \, .
\end{equation}
The quantity $q$ is also known as the \textit{loss cone occupation fraction} or \textit{loss cone diffusivity}, since the local strength of stochastic kicks determines how many particles on an orbit penetrating $r_\mathrm{TDE}$ will actually reach the pericentre without being scattered back onto a safe orbit. Consequently, inside the loss cone,  no particles with $\R < \R_\mathrm{TDE}$ are statistically expected when $q \ll 1$ (since relaxation needs more than one orbit to change significantly the pericentre), while many particles are expected at any time when $q\gg1$ (as their pericentre is likely to change along the orbit due to 2-body encounters).
The former is usually referred to as the \textit{empty loss cone regime}, the latter as the \textit{full loss cone regime}. The number of orbits required for relaxation to affect the orbital parameters will be a crucial point for our analysis of repeated partial disruptions, since they can happen only if the orbital parameters are negligibly altered by relaxation.
Because of the dependency on $r_\mathrm{TDE}$, the value of $q$ in the two models is related as
\begin{equation}
    q^B(E) = 1.8 \; q^A(E) \, .
\end{equation}

A common assumption in literature is to employ a reduced equation to evolve a distribution function $f(E)$ that does not depend explicitly on $\R$, and assumes the steady state distribution of angular momentum at each energy
\begin{equation}\label{eq:CK-R0}
    f(E,\R) = f_0 \, \ln \frac{\R}{\R_0}; \quad \ln \frac{\R_0}{\R_\mathrm{TDE}} = -\sqrt[4]{q^2+q^4}
\end{equation}
to compute the rate of tidal disruptions at each energy. The relation between $\R_0$ and $\R_\mathrm{TDE}$ is directly determined by the model of instantaneous tidal disruptions at the pericentre. The rates are computed from the solution $f(E)$ of the 1D evolution in energy \citep{vasilievNewFokkerPlanckApproach2017, Stone2020}
\begin{equation}
    f(E) = f_0 \, \int_{\R_\mathrm{TDE}}^1 d \R \ln \R / \R_0
\end{equation}
as
\begin{equation}
    \frac{d N_\mathrm{TDE}}{dt} =  \int\, dE \; 4 \pi^2 \, J_c^2\,P \, D \, f_0
\end{equation}
where the pre-factor $4 \pi^2 \, J_c^2\,P$ converts the distribution function $f$ into the differential distribution $n(E)$ such that the number of stars is given by $\int dE \, n(E)$. As we shall see, the model of a TDE as an instantaneous event is directly related to the estimate of TDE event rates through the computation of $f_0$.

We summarise here the key assumptions made to derive the relaxed profile in \citet{cohnStellarDistributionBlack1978}:
\begin{enumerate}[CK 1]
    \item A full, terminal disruption occurs at the pericentre if and only if $r_\mathrm{p} < r_\mathrm{TDE}$.
    \item\label{item:CK2} The energy fluctuation timescale is much longer than the angular momentum relaxation timescale.
\end{enumerate}

\begin{figure}
    \includegraphics[width=\linewidth]{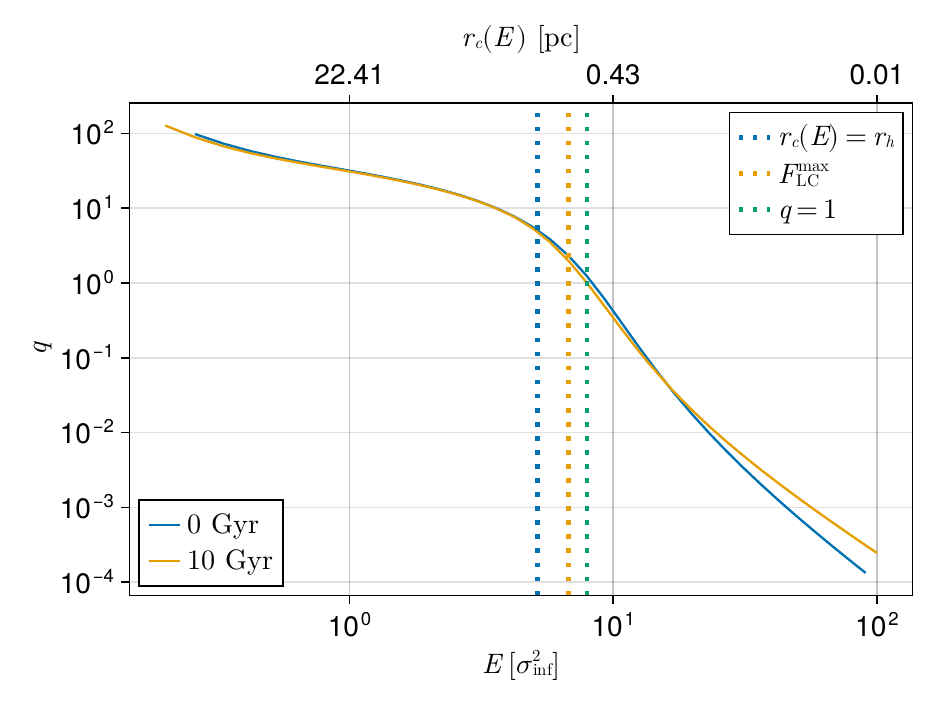}
    \caption[Loss cone diffusivity as a function of energy]{\label{fig:q_as_a_function_of_E} The value of the loss cone diffusivity $q$ as a function of energy (in units of the velocity dispersion at the influence radius) in a system with a central massive black hole of mass $M_\bullet=4 \times 10^6M_\odot$ and for a sun-like star. The stellar distribution is composed of Solar mass stars and stellar mass black holes, both following a Dehnen profile with inner slope $\rho \propto r^{-1.5}$. We show the curve $q(E)$ both for the initial conditions (blue) and after $10$ Gyr of relaxation (orange). The upper axis shows the radius of the circular orbit at energy $E$.  The details of the underlying distributions of stars are in the text. We mark with vertical dotted lines from left to right: in blue, the energy corresponding to a circular orbit at the influence radius; in orange, the energy where the flux entering the loss cone is maximised; and in green, the energy where $q=1$.}
\end{figure}

\begin{figure*}\centering  \includegraphics[width=0.49\linewidth]{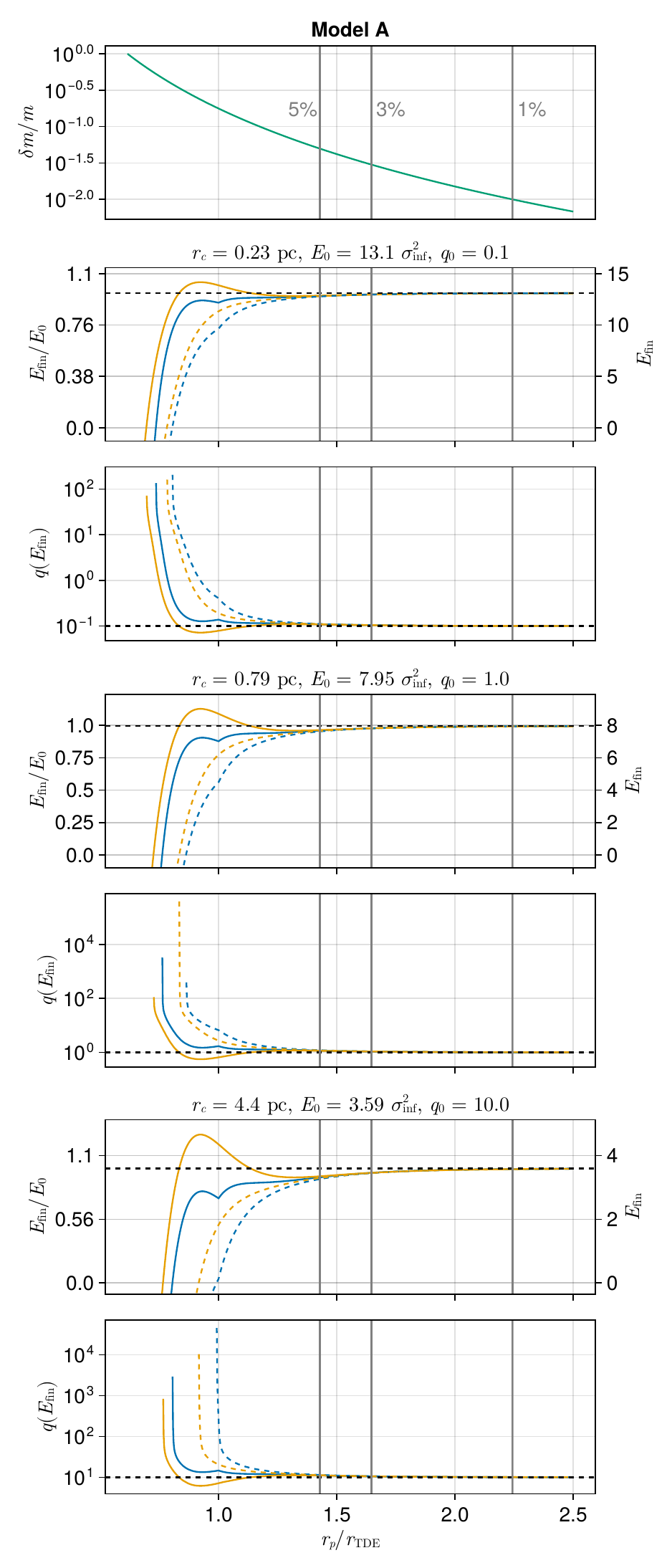}\includegraphics[width=0.49\linewidth]{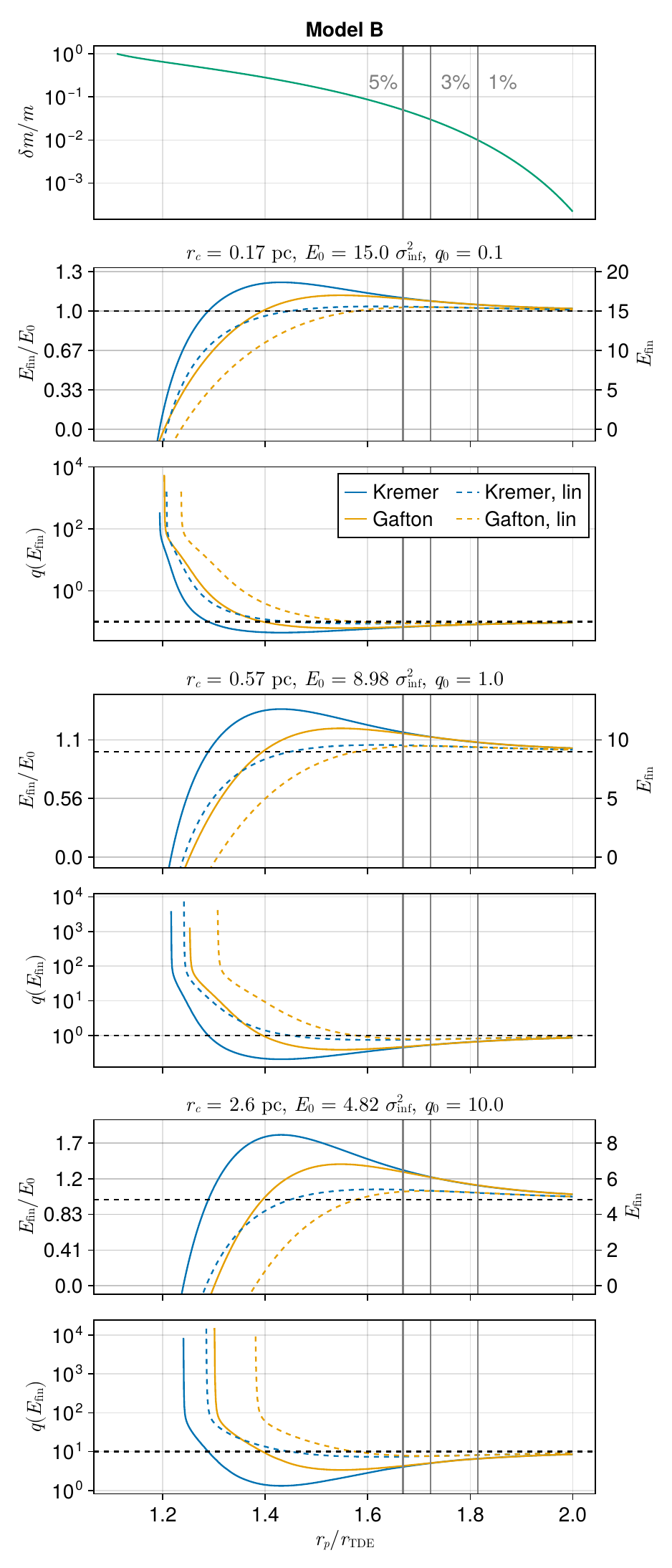}\\
    \caption[Effects of a single partial TDE in the empty, intermediate, and full loss cone regime.]{\label{fig:single_passage}Summary of the effects of mass-loss and tidal excitation for a single passage of a partial disruption.  Different sets of rows represent stars with different initial energies, lying in the empty, intermediate, and full loss cone regimes. On the left, we consider model A ($m_\star = 1\, M_\odot$, $n = 3.0$), and on the right, model B ($m_\star = 0.3 M_\odot$, $n = 1.5$).
    In each column, we plot in the first row the relative mass loss, as a function of the stellar pericentre divided by the tidal radius. Then we show three figures (corresponding to $q_0=\{0.1,\ 1.0,\ 10.0\}$) composed of two panels each. In the first panel we show the energy after the disruption $E_\mathrm{fin}$ through its ratio to the initial energy $E_0$ according to the model for core kicks  from Eq. 10 (blue) and Eq. 11 (orange).  For each core kick model, we consider both non-linear prescriptions for tidal excitation (solid) and the linear regime (dashed). In the second panel of each figure we show the loss cone filling factor for the final energy $q(E_\mathrm{fin})$. Note that $q$ goes to infinity as $E_\mathrm{fin}\to0$, since the star gets unbound. In all plots we show vertical dashed lines at the pericentres corresponding to $\{1\%,\ 3\%,\ 5\%\}$ mass stripping, and horizontal dashed lines corresponding to the initial energy $E_0$ and the initial diffusivity $q(E_0)$. Due to the different stellar structures, $\delta m (r_\mathrm{p})$ has different trends at large $r_p$ for model A and model B. We see that at large pericentres - where mass loss is small - stars described by model A receive small negative kicks and are moved to less bound orbits, while stars described by model B are moved to slightly more bound orbits.}
\end{figure*}
\begin{figure*}
	\centering
    \includegraphics[width=0.8\linewidth]{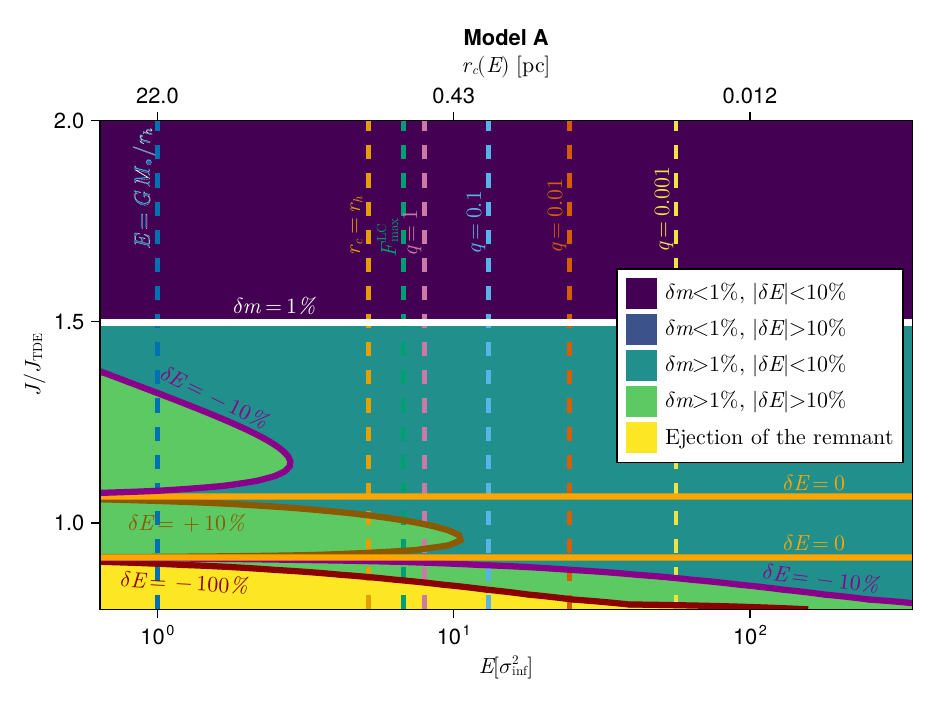}\\
    \includegraphics[width=0.8\linewidth]{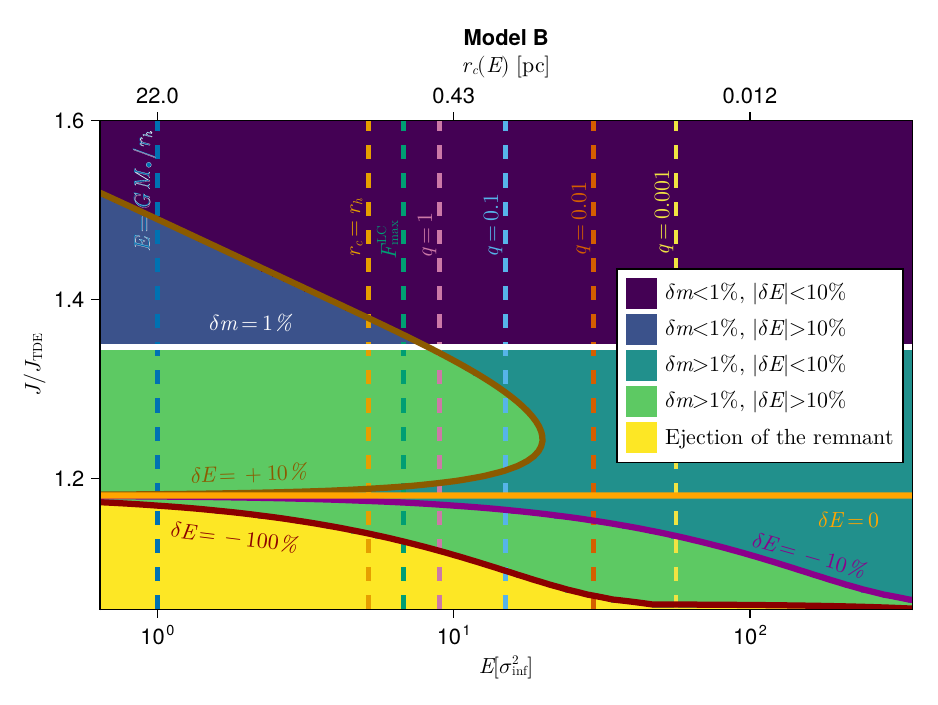}
    \caption[Overview of the effect of a single partial TDE in a galactic nucleus]{\label{fig:overview} Overview of the effect of a single pericentre passage due to mass-loss  and tidal excitation in phase space, where we express the angular momentum in units of $J_\mathrm{TDE}$ (the orbit with pericentre $r_\mathrm{p}$) and energy in units of the squared velocity dispersion at the influence radius $\sigma^2_\mathrm{inf}$. In the upper figure we consider a star described by model A, in the lower figure we consider a star described by model B. We consider Eq.~\ref{eq:mass_loss_gafton} to account for mass loss effects, and the non-linear model for tidal excitation of
    Eq.~\ref{eq:nonlin_tid}. In each case we identify here the regions corresponding to different effects on the orbit (defined arbitrarily): negligible effects ($\delta m < 1\%,\ \delta E/E<10\%$), significant change in energy ($|\delta E/E| > 10\%$ disregarding the mass change) and ejections ($\delta E/E = -100\%$). We report some reference values of $E$ as dash vertical lines, and the upper axis shows the corresponding radius of the circular orbit $r_c$ with energy $E$ (including the stellar potential).}
\end{figure*}

We will now consider a prototypical system to compute the strength of 2-body relaxation. The system is composed of a central MBH with mass $M_\bullet=4\times10^6 M_\odot$ surrounded by a distribution of stars (total number $8 \times 10^7$, individual mass $1 M_\odot$) and stellar black holes (total number $8 \times 10^4$, individual mass $10 M_\odot$) both following a Dehnen profile \citep{dehnenFamilyPotentialDensityPairs1993, tremaineFamilyModelsSpherical1994} with scale radius of $8.8\ \mathrm{pc}$ and inner slope $1.5$, as in \cite{broggiExtremeMassRatio2022}.
This system has an influence radius
\footnote{There are multiple definitions of influence radius in literature. In this work we define $r_\mathrm{inf} = G\, M_\bullet / \sigma^2_\mathrm{inf}$, where $\sigma_\mathrm{inf}$ is the local velocity dispersion.}
of $r_\mathrm{inf} \simeq 2.2\  \mathrm{pc}$ and the velocity dispersion at the influence radius corresponds to $\sigma_\mathrm{inf}=88\ \mathrm{km/s}$, computed according to the $M-\sigma$ by \citet{gultekinTHEMUpsigmaANDMLRELATIONS2009}.
This system is used to compute the quantity $\left\rvert\Delta \R\right\rvert_{orb}$.
The loss cone diffusivity $q = q(E)$ for the initially isotropic distribution and for the final profile (at $10$ Gyr) in this system is shown in Fig.~\ref{fig:q_as_a_function_of_E}.
The net effect of mass segregation, which pushes stars to lower $E$ and stellar black holes (sBHs) towards higher $E$, is to increase $q$ for orbits with $a \simeq r_c(E) \lesssim 10^{-1}$ pc. In fact, in this region $m_\star \, \rho_\star \leq  m_\mathrm{sBH}\, \rho_\mathrm{sBH}$ and therefore stellar mass black holes dominate two-body relaxation rates. In the rest of this work, we use the estimate of $q$ from the relaxed profile at $10$ Gyr.

In Fig.~\ref{fig:single_passage} we show the predictions of our semi-analytical model for partial disruptions of a non-oscillating star. We consider both model A and model B stars; different values of the initial energy $E_0$ corresponding to $q=\{0.1, 1.0, 10.0\}$; different prescriptions for energy change in the core kick (Eqs.~\ref{eq:mass_loss_Kremer} and \ref{eq:mass_loss_gafton}); and different models for tidal excitation (linear and non-linear).

Irrespective of the specific models used for the computation of energy changes (i.e. tidal excitation and mass loss), there is a qualitative difference between the partial disruption of stars in the two stellar models driven by the different internal structure.
The structure of model $B$ requires deeper penetration of the orbit (smaller $r_\mathrm{p}$) for the star to lose mass, so that the function $\delta m / m$ goes to zero at large pericentres faster than to stars described by model A. Moreover, model B corresponds to a larger energy stored into oscillations, with tidal oscillations being dominant at large $r_\mathrm{p}$.
On the other hand, partial disruptions of model $A$ stars are dominated by the effects of mass loss at larger pericentres. In general, the linear expansion seems to perform well compared to non-linear estimates when mass loss is below approximately $5\%$. While in the case of model A Eq.~\ref{eq:mass_loss_gafton} predicts a larger energy increase with respect to Eq.~\ref{eq:mass_loss_Kremer}, the opposite is true in the case of model B given the larger pericentres involved in units of $r_\mathrm{TDE}$.

To overview the full energy range, in Fig.~\ref{fig:overview} we show the qualitative effect of single passages pf a non-oscillating star at the pericentre as a function of the initial energy $E$ and angular momentum $J$ for model A and model B; we also report the radius $r_c$ of a circular orbit at energy $E$ for better interpretation.
Here, we use the combination of energy change given by Eq.~\ref{eq:mass_loss_gafton} and non-linear tidal excitations of Eq.~\ref{eq:nonlin_tid}.
We plot the contours that correspond to $\Delta E/E = -100 \%$, $|\Delta E/E| = 10\%$ and $\delta m / m < 1\%$, and define some (arbitrary) thresholds to classify the possible outcomes:
\begin{itemize}
    \item \textbf{Negligible change in the orbit, negligible change in the mass}. We define this to happen when $|\Delta E|/E < 10 \%$ and $|\delta m| / m < 1\%$ (dark blue).
    \item \textbf{Significant change in the orbit, negligible change in the mass.} This happens when $ 10 \% < |\Delta E|/E < 100\%$ and $|\delta m|/m < 1\%$ (blue). This only happens in model B.
    \item \textbf{Negligible change in the orbit, significant change in the mass.} This happens when  $|\Delta E|/E < 10 \%$ and $|\delta m|/m > 1\%$ (aqua green).
    \item \textbf{Significant change in the orbit, significant change in the mass.} This happens when $ 10 \% < |\Delta E|/E < 100\%$ and $|\delta m|/m > 1\%$ (green).
    \item \textbf{Ejection of the remnant.} This happens when $\delta E < -100\%$ (yellow).
\end{itemize}

The level curves of model A are non-monotonic, and for $E \lesssim 3 \, \sigma^2_\mathrm{inf}$ there are two disconnected curves corresponding to $\delta E = - 10 \%$.
This is consistent with the trend of $E_\mathrm{fin}/E_0$ shown in Fig.~\ref{fig:single_passage}: for model A, this ratio has a maximum (with $E_\mathrm{fin} / E_0 > 1$) and a minimum (with $E_\mathrm{fin} / E_0 < 1$ ). The structure of the levels $\delta E = \pm 10\%$, moving from small to large pericentres, is the following: \textit{(i)} for any value of $E_0$ there is a small value of $r_p$ such that $E_\mathrm{fin} = 0.9 E_0$ and $\delta E = -10\%$; \textit{(ii)} if the maximum of $E_\mathrm{fin}$ is larger than $1.1 E_0$, there are two values of $r_p$ such that $E_\mathrm{fin} = 1.1 E_0$ and $\delta E = +10\%$; \textit{(iii)}
whenever the local minimum of $E_\mathrm{fin}$ is below $0.9 \, E_0$, there are other two values of $r_p$ such that $E_\mathrm{fin} = 0.9 E_0$ and $\delta E = -10\%$.
The level curves of model B have a simpler pattern, since $E_\mathrm{fin}/E_0$ only has a maximum with $E_\mathrm{fin} / E_0 > 1$ and is asymptotic from above to $1$ at large $r_p$. Therefore, the structure of the levels $\delta E = \pm 10\%$, moving from small to large pericentres, is the following: \textit{(i)} for any value of $E_0$ there is a small value of $r_p$ such that $E_\mathrm{fin} = 0.9 E_0$ and $\delta E = -10\%$;  \textit{(ii)} if the maximum of $E_\mathrm{fin}$ is larger than $1.1 E_0$, there are two values of $r_p$ such that $E_\mathrm{fin} = 1.1 E_0$ and $\delta E = +10\%$.

Overall, when $\delta m < 1\%$ model A stars have smaller energy fluctuations, while the tidal excitation is strongly relevant in model B for semimajor axis (estimated through $r_c$) comparable to the influence radius. On the other hand, mass loss effects are dominant for model A and the total energy change is positive (i.e, the resulting orbit is more bound) only in an interval of values at small pericentres. Close to $r_\mathrm{p}/r_\mathrm{TDE} = \eta_\mathrm{TTDE}$, the energy fluctuation steeply decreases and becomes negative, so that the remnant is ejected in both models.

\section{Repeated disruptions and 2-body relaxation}
We will now try to assess the impact of single and repeated disruptions, first considering the full loss cone regime, where repeated disruptions are unlikely to occur, and then the empty loss cone regime where repeated partial disruptions are likely to take place.
The key variable we will introduce is the number of orbits needed by relaxation to significantly alter the pericentre
\begin{equation}
    N_q \simeq \frac{\R}{\left\langle \Delta \R \right\rangle_\mathrm{orb}} = \frac{1}{q}
    \label{eq:npassages}
\end{equation}
so that a repeated partial disruption is likely to occur when $N_q \gtrsim 2$.

\subsection{Full loss cone regime}

When $q\gtrsim1$, relaxation acts on timescales shorter than times between two partial disruption events. In this regime, single passages grazing or entering the loss cone radius will alter the orbital parameters once per period $P$, while the timescale of relaxation in angular momentum is $P/q$.
This implies that only a small fraction of the stars instantaneously moving on an orbit with pericentre smaller than the tidal disruption radius will actually manage to reach such a pericentre, but they are likely to be scattered to a different (safe) orbit before reaching it.
As a consequence, the probability of having a pTDE and staying on the same orbit to undergo a second disruption is small, and virtually negligible since $q$ grows quickly as the orbital energy $E$ decreases (see Fig.~\ref{fig:q_as_a_function_of_E}).

A fraction of partial disruptions of loosely bound stars may result in the ejection of the remnant at a pericentre larger than $r_\mathrm{TDE}$ (see Fig.~\ref{fig:overview}), and this may be compatible with classical loss cone theory.
In fact, since they both imply an instantaneous removal of the subject star from the distribution, ejections can be treated together with total disruptions in the standard approach to the loss cone, where particles are removed when they reach a pericentre below the critical value $r_\mathrm{TDE}$. Therefore, by identifying a suitable, larger threshold radius $r_\LC > r_\mathrm{TDE}$ that roughly distinguishes the region of ejections plus total disruptions, classical loss cone theory may be simply adapted to provide a good description of the system. Such a value of $r_\LC$ weakly depends on energy, but this does not affect the estimation of the corresponding Cohn-Kulsrud solution.

\subsection{Empty loss cone regime}\label{subsec:repeated}
In the empty loss cone regime $q \ll 1$, and repeated partial disruptions are expected. In fact, the empty loss cone regime is defined as the region of phase space where relaxation needs many orbits to alter significantly the orbital parameters of an orbit with the critical pericentre $r_\mathrm{TDE}$.
\begin{figure*}
    \centering
        \hspace{0.1\linewidth}\textbf{Model A} \hspace{0.4\linewidth} \textbf{Model B}\\
        \includegraphics[width=0.49\linewidth]{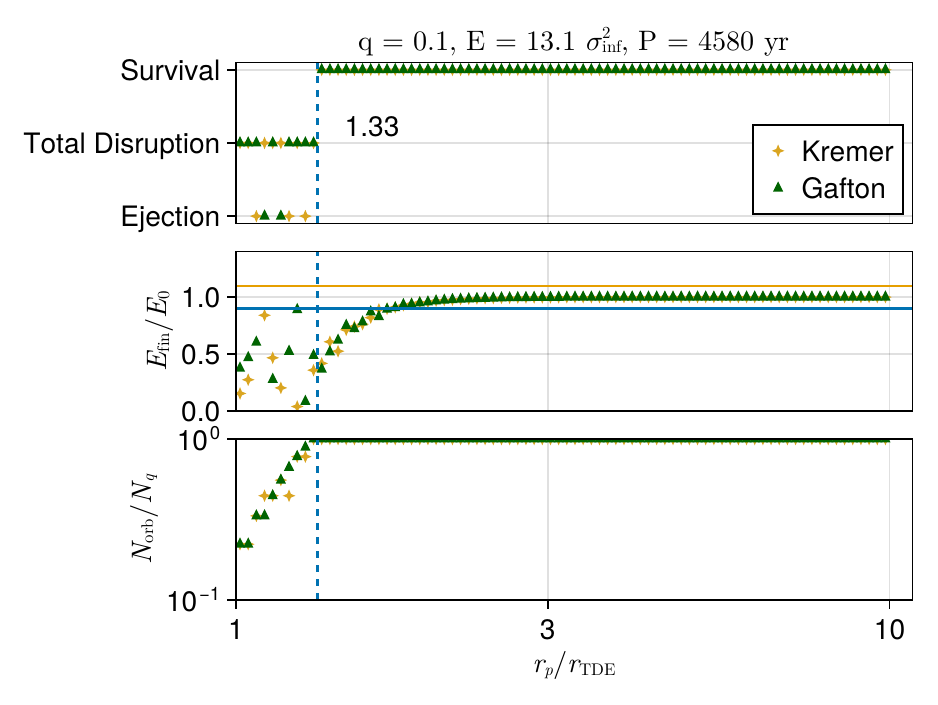}
        \includegraphics[width=0.49\linewidth]{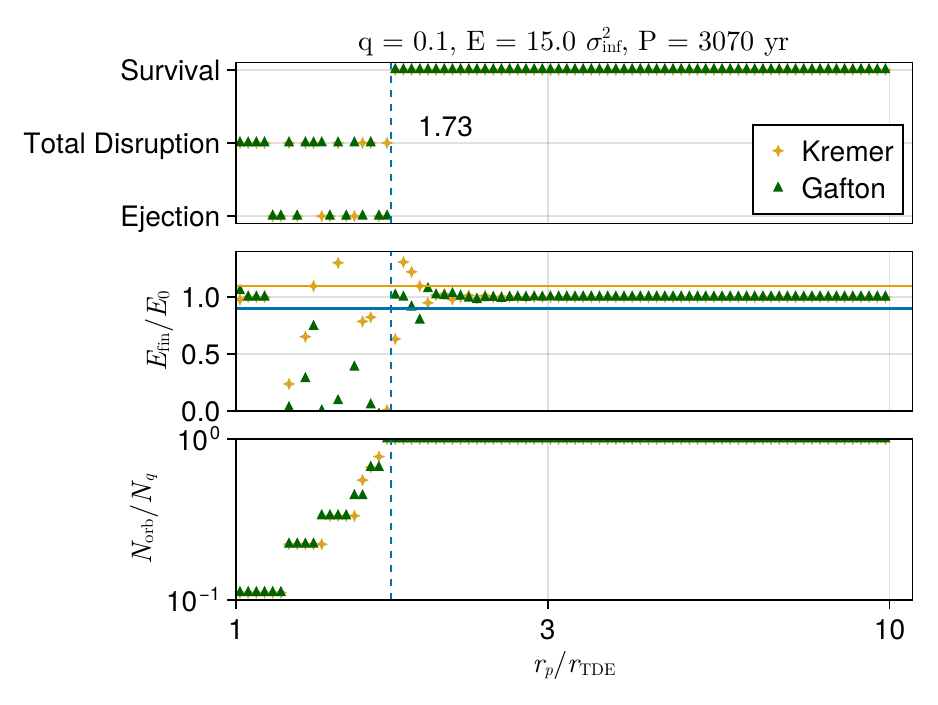}\\
        \includegraphics[width=0.49\linewidth]{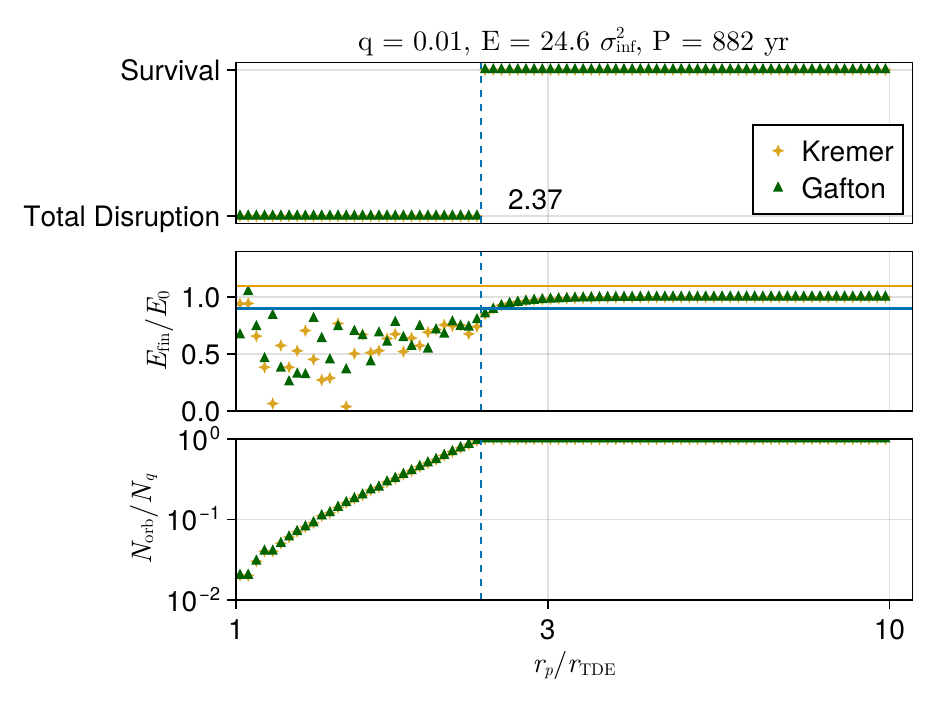}
        \includegraphics[width=0.49\linewidth]{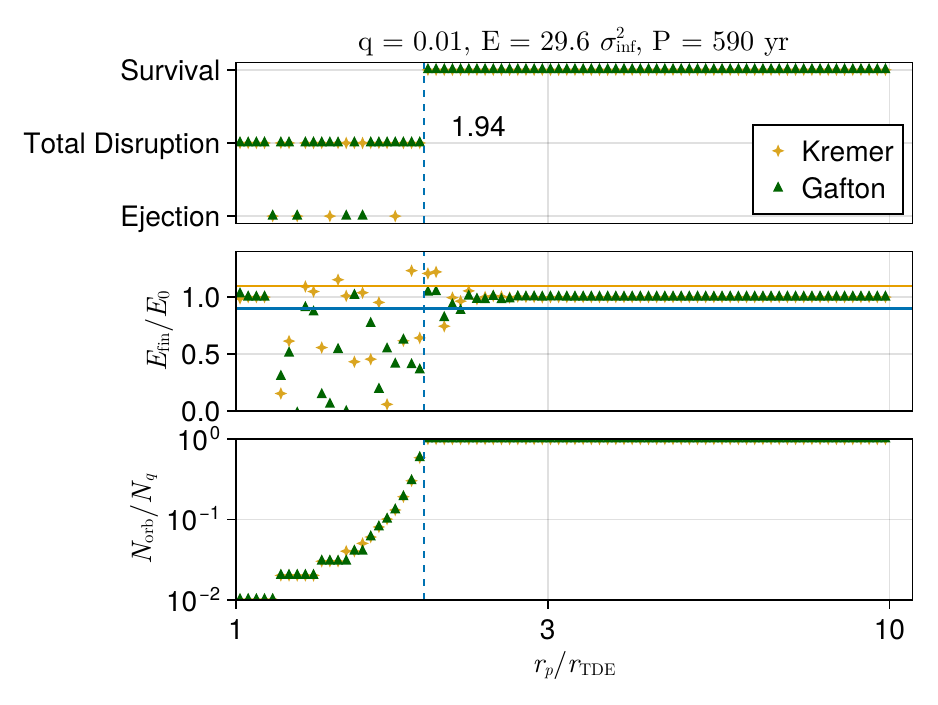}\\
        \includegraphics[width=0.49\linewidth]{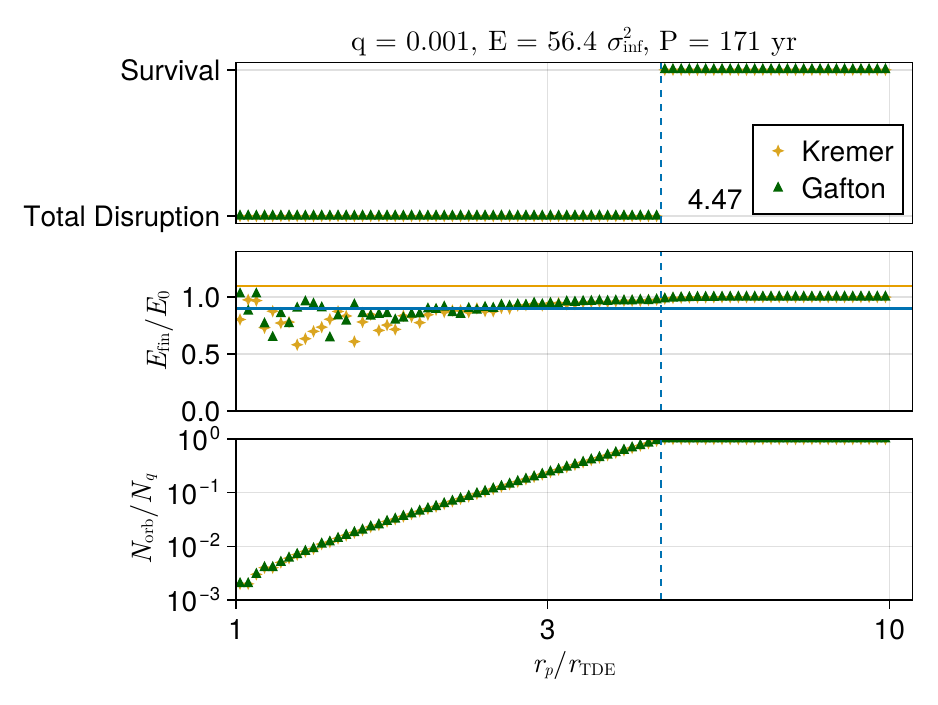}
        \includegraphics[width=0.49\linewidth]{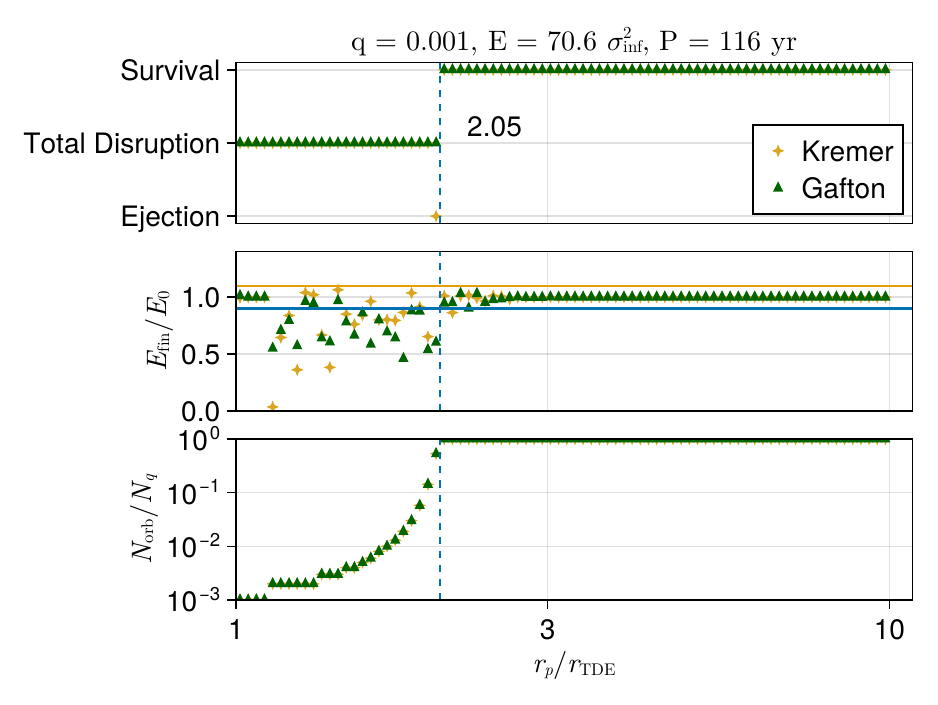}
        \caption[Effect of repeated pTDEs in the empty loss cone regime at different values of the loss cone diffusivity.]{
        \label{fig:q_and_pTDE} Effects of repeated partial disruptions with initial pericentre $r_\mathrm{p}$ (in units of $r_\mathrm{TDE}$) according to our simple step-by-step evolution for $q=0.1$ (upper figure), $q=0.01$ (middle figure) and $q=0.001$ (lower figure); the corresponding energy for our fiducial model is also reported in units of $\sigma^2_\mathrm{inf}$ and the orbital period $P$ in years. The first column refers to model A stars, the second column to model B stars. In the upper panel of each figure, we show the outcome of particles completing less than $N_q = 1/q$ orbits, distinguishing between total disruptions ($r_\mathrm{p} < r_\mathrm{TDE}$) and ejections ($E_\mathrm{fin}<0$); we dub as survival the stars that complete $N_q$ orbits. In the middle panel, we report the ratio between the final and initial energy, showing the $90\%$ line (blue) and the $110\%$ line (orange) for reference. In the lower panel, we show the fraction of completed orbits $N_\mathrm{orb}/N_q$. We mark with a vertical dashed line the threshold pericentre, i.e. the smallest pericentre allowing the stars to complete $N_q$ orbits. Model A stars show a reduced orbital energy as they start closer to the threshold pericentre, meaning that they are less bound to the central MBH. On the other hand, model B stars show significant energy fluctuations slightly biased towards less bound orbits when computing the effect of asymmetric mass loss according to \citet{gaftonRelativisticEffectsTidal2015}.}
\end{figure*}

Since we are interested in understanding whether the classical loss cone theory applies to partial disruptions, we need to assess the cumulative effect of repeated partial disruption on the orbits over the local relaxation timescale, that operates in $N_q$ orbits; furthermore, this approach allows us to estimate how many passages a star initially on a partial disruption orbit manages to complete before either being ejected, surviving the interaction without escaping the system, or being fully disrupted.

For each pericentre passage, we will consider the mass lost according to Eq.~\eqref{eq:mass_loss_ryu} (model A) or Eq.~\eqref{eq:mass_loss_guillochon} (model B) and the two prescriptions for the corresponding decrease in specific binding energy via core kicks, also including the exchange of energy between stellar oscillations and orbital motion according to Eq.~\eqref{eq:nonlin_tid}. We remark that the perturbation due to tidal excitation is positive-definite (i.e. the orbit becomes more bound) only at the first pericentre passage.
At subsequent passages, tidal deformation will inject or subtract energy to the stellar orbit depending on the phase of the ongoing oscillation (induced by the previous passage) when the star returns to pericentre.
To include this effect, we use Eq.~\ref{eq:nonlin_tid} to determine the amplitude of the energy exchanged but, from the second pericentre passage onwards, we give it a random sign at each passage.

For $q = \{0.1, 0.01, 0.001\}$ and each stellar model we set up three simulations with the following simple scheme.
\begin{enumerate}
    \item \label{itm:j_zero} We set the initial energy $E_0$ corresponding to the chosen value of $q$. We set the initial mass to $m_\star$ and the initial angular momentum $J_0 \in [J_\mathrm{TDE}, \sqrt{5} \, J_\mathrm{TDE}]$, where $J_\mathrm{TDE}$ is the angular momentum of the orbit whose pericentre is $r_\mathrm{TDE}$.
    \item We compute the pericentre $r_\mathrm{p}$ and use Eq.~\eqref{eq:mass_loss_ryu} or Eq.~\ref{eq:mass_loss_guillochon} to determine the mass lost at the 
    pericentre passage.
    \item We use Eq.~\eqref{eq:mass_loss_Kremer} or Eq.~\eqref{eq:mass_loss_gafton} and Eq.~\eqref{eq:nonlin_tid} to compute the total energy variation
    \begin{equation}
        \delta E = \delta E_\mathrm{ml} + \xi \, \delta E_\mathrm{tid}
    \end{equation}
    where $\xi = 1$ at the first pericentre passage, and $\xi = \pm 1$ with equal probability at later passages.
    We keep track of the total energy of the oscillations, and we adjust $\delta E_\mathrm{tid}$ to damp them completely when their total energy would become negative.
    \item We compute the new mass $m_\star$, the new tidal disruption radius $r_\mathrm{TDE}$, and the new energy $E$.
    \end{enumerate}
We repeat steps $2-4$ until
\begin{itemize}
 \item \textbf{total disruption} ($r_\mathrm{p} < r_\mathrm{TDE}$ for model A, $r_\mathrm{p} < \eta^B_\mathrm{TTDE}\,r_\mathrm{TDE}$ for model B),
 \item  \textbf{ejection} ($E < 0$),
 \item \textbf{survival} (i.e. the star completes $N_q$ orbits).
\end{itemize}
For each value of $q$ we choose 80 equally spaced values in $\ln J$
\begin{equation}
    \ln J_i = x_0 + i \, \delta \qquad i=0 \dots 79
\end{equation}
with $x_0 = \ln \sqrt{10} /160$ and $\delta = \ln \sqrt{10}/80$, consistently with step \ref{itm:j_zero} above.

We treat $R_\star$ and $\eta_\mathrm{TTDE}$ as fixed parameters throughout the evolution, since modelling the readjustment of the star requires to directly solve the complex hydrodynamic evolution of the stellar structure\footnote{Since we are considering orbital periods of at most $10^4$ yr,
much less than the Kelvin-Helmholtz timescale \citep[$\sim 1-100$ Myr;][]{kippenhahnStellarStructureEvolution1990}, dramatic mass loss and energy injection through the dissipation of nonlinear oscillation modes can in principle seriously affect the internal structure of the star.}. In Appendix \ref{ap:therm} we explore a non-fiducial scenario where the star can inflate due to nonlinear tides. In this case, part of the energy stored in oscillations is converted into internal energy through an enlargement of the radius. Finally, we remark that the procedure composed of steps $1-4$ implies that the specific angular momentum is constant throughout the repeated disruptions.

In Fig.~\ref{fig:q_and_pTDE} we show
the results of our simulations by reporting the outcome (survival, total disruption or ejection), the final energy and the number of repeating partial disruption orbits for each simulation as a function of its initial pericentre $r_\mathrm{p}$. For both models and for any value of $q$, we see that repeating the partial disruption process can lead to total disruption and, in a minority of cases, to ejection at pericentres larger than $r_\mathrm{TDE}$ and $\eta_\mathrm{TDE} \, r_\mathrm{TDE}$.
The threshold radius (reported in units of $r_\mathrm{TDE}$ in each sub-figure) increases with $N_q$ and therefore decreases with $q$.
There are, however, some significant differences between the repeated partial disruption of a star described by model A and one described by model B.
\begin{figure*}
 \includegraphics[width=0.49\linewidth]{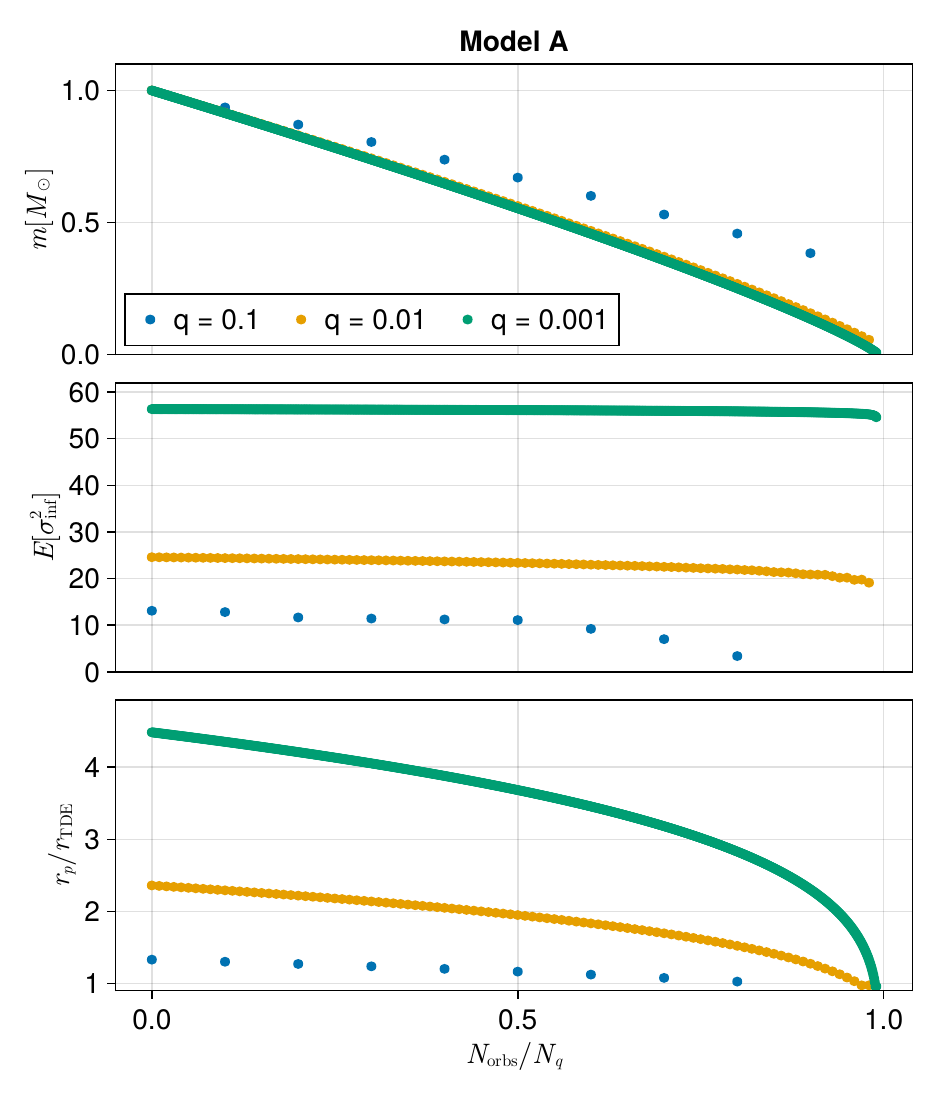}
 \includegraphics[width=0.49\linewidth]{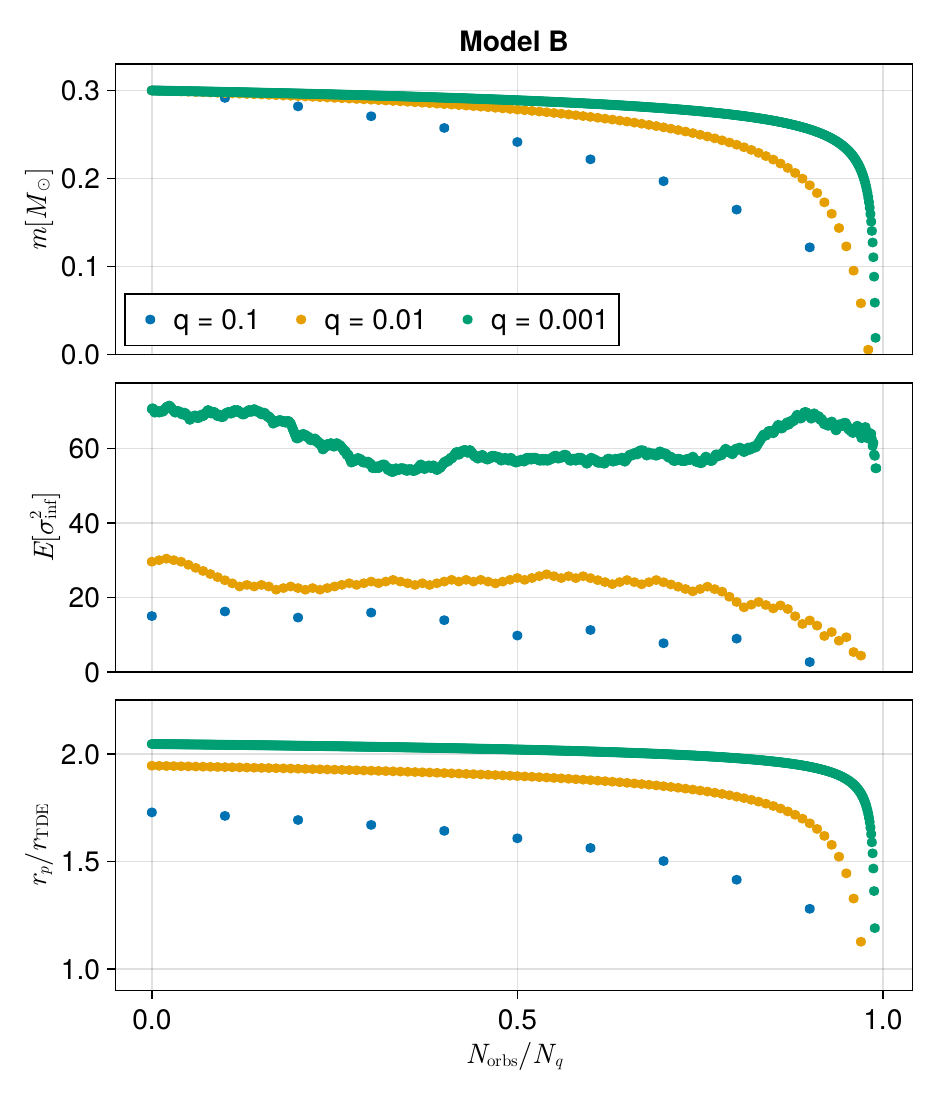}
\caption{\label{fig:threshold_orbits} Energy, mass and pericentre in units of $r_\mathrm{TDE}$ for the orbits corresponding to the threshold values highlighted in Fig.~\ref{fig:q_and_pTDE}. As $q$ decreases, the mass of the star before the (final) total disruption decreases, possibly resulting in less energetic events.}
\end{figure*}

In the case of model A, when a star is partially destroyed and manages to survive $N_q$ pericentre passages, it gradually decreases its specific binding energy $E$, ending up on orbits for which $q$ is larger: particles slightly above the threshold radius at $q^A(E_0)$ equal to $0.1$ or $0.01$ will decrease $E$ by $\sim 15\%$ and $\sim 50\%$ respectively.
This is consistent with the fact that asymptotically the total energy fluctuation is dominated by the effects of mass loss.

On the other hand, in the case of model B the star may increase its energy up to $\sim10\%$ for the three values of $q$ we considered.
This is consistent with the fact that asymptotically the total energy fluctuation is dominated by the effects of tidal excitation.

We highlight that, for the cases we considered, each threshold pericentre between total disruption / ejection and survival is consistent between the two models for core kicks, and is such that the linear regime of tidal oscillations is sufficiently accurate (see Fig.~\ref{fig:single_passage}). Only below the threshold pericentre the non-linear description of tidal oscillations must be taken into account.

In Fig.~\ref{fig:threshold_orbits} we show the trajectories at the threshold pericentre for both models. The plot shows that model A corresponds to a gradual decrease of $E$, while model B corresponds to fluctuations of $E$ because of tidal excitation.
By construction, the angular momentum of the star is conserved along the evolution, so that the pericentre is constant too: $r_\mathrm{p}$ is recomputed at each step with the updated energy, but the fluctuations are at most of one part in $10^5$ for these orbits.
Therefore, the key change in the process is the enlargement of $r_\mathrm{TDE}$ due to mass loss, so that within a small number of orbits (compared to $N_q$) the mass is fast depleted up to the total disruption (upper panel).
The energy of the star, on the other hand, shows an evolution that depends on the stellar model and the particular pericentre, and for $q=0.01$ and $q=0.001$ clearly follows the expectation based on the large $r_\mathrm{p}/r_\mathrm{TDE}$ trend.

\section{Discussion and conclusions}
We presented a simple semi-analytical model of partial disruptions for two stellar models, one for main sequence stars with mass $1 \, M_\odot$ and the other for main sequence stars with mass $0.3 \, M_\odot$. The prescriptions included in our semi-analytical models are based on the results of hydrodynamical simulations and semi-analytic modeling \citep{Ryu2020a,Ryu2020b,Ryu2020c, Ryu2020d, guillochonHydrodynamicalSimulationsDetermine2013, gaftonRelativisticEffectsTidal2015, kremerHydrodynamicsCollisionsClose2022, stoneConsequencesStrongCompression2013, ivanovNewModelTidally2001}.
To properly study the phenomenon in the empty loss cone regime, where relaxation needs more than one orbit to change the orbital parameters, we accounted for the possibility of repeating partial disruptions. We will now discuss the implications of our model for the inclusion of partial disruptions in the loss cone theory, and the computation of their event rate.

\subsection{Partial disruptions and loss cone theory}
In the theory of the loss cone, one can use a Monte Carlo or a Fokker Planck approach to study the evolution of the distribution function of a stellar system \citep{cohnStellarDistributionBlack1978, shapiroStarClustersContaining1978}. These studies show that particles enter the critical radius of the loss cone once are driven there by the stochastic encounters; in particular they typically start with a pericentre larger than the critical value and are then driven on orbits penetrating the critical sphere.

In the full loss cone regime, particles approaching the loss cone boundary can be scattered at any pericentre value \citep{merrittDynamicsEvolutionGalactic2013}, so that all possible outcomes (i.e. partial disruption with orbital and mass change, ejection or total disruption) can occur. Neglecting the change in energy and mass, one can apply the classical loss cone theory - based on the instantaneous disruption model - to account for both ejections and disruptions. The standard treatment should be applied with the replacement of the tidal disruption radius $r_\mathrm{TDE}$ with the largest pericentre where ejections appear (as a function of energy). The critical pericentre corresponding to ejections is generally larger than $r_\mathrm{TDE}$, as can be seen in Fig.~\ref{fig:overview}.

In the empty loss cone regime, the stellar model plays an important role. For stars of $1 M_\odot$ (i.e. star model A), our model predicts that for $q\simeq 0.01$ (and larger values up to $q \sim 1$) stars approaching the loss cone in phase space by gradually reducing their pericentre due to 2-body interactions will likely \textit{reduce} their specific binding energy significantly (more than $10\%$). Consequently, stars will move to a region where $q$ is larger - where they can be scattered away from the loss cone, or being disrupted / ejected at an energy different from the initial one. Only at very small $q$, as in the case of $q=0.001$, the energy change is so small that the particle will likely proceed towards total disruption through a sequence of very small partial disruption events. The classical picture based on instantaneous disruption cannot be therefore adapted, as we proposed for the full loss cone regime. Here, the phenomenon of partial disruptions will effectively alter the energy of the particles approaching the loss cone, invalidating the assumption CK\ref{item:CK2}, i.e. the fact that the (stochastic) relaxation of $\R$ can determine the expected relaxed profile in phase space.

Stars described by model B, on the other hand, show a difference: the specific binding energy is subject to significant fluctuations, as these compact stars are more susceptible to tidal excitation. The orbital energy can change by more than $10\%$ as the star approaches the loss cone boundary, with a bias towards tighter configurations when considering Kremer's estimate for mass loss (Eq.~\ref{eq:mass_loss_Kremer}). This means that the star may be moved to regions where $q$ is smaller and two-body interactions less efficient. This process is somehow similar to that of extreme mass ratio inspirals (EMRIs), a phenomenon where a compact object approaches the central MBH so close that its evolution is deterministically driven by GWs emission over hundreds of thousand of orbits \citep{amaro-seoaneRelativisticDynamicsExtreme2018}. In this case, the role of the emission of GWs at each pericentre passage would be replaced by a small mass loss and consequent tidal excitation. The total event rate of EMRIs (but not the distribution at capture) is computed by integrating the classically expected rate -- with the loss cone radius equal to the innermost parabolic orbit -- over the region where one expects EMRIs to be formed \citep{hopmanOrbitalStatisticsStellar2005, bar-orSteadyStateRelativistic2016}. For stars of model B, a tentative approach to compute the total number of stars undergoing a partial disruption may be therefore to integrate the classical rate over the phase-space region where $N_q \gtrsim 2$. However, the information about the number of repeated partial disruptions that each of these stars undergoes must be inserted with a more detailed model.

As we mentioned at the beginning of this section, relaxation in the empty loss cone regime mainly drives particles from larger to smaller angular momenta/pericentres. This means that stars will gradually approach the critical thresholds identified in Fig.~\ref{fig:q_and_pTDE} from right to left in the plots, effectively resulting in a sequence of weak partial disruptions as shown in Fig.~\ref{fig:threshold_orbits}. As $q$ decreases, the mass lost at each pericentre passage is smaller, and therefore the resulting electromagnetic flare is expected to be weaker \citep{Stone2020, rossiProcessStellarTidal2021}.

\subsection{Implications for TDE rates}
Currently available estimates of time-evolving TDE rates rely on the assumption that the relaxed profile in angular momentum of stars is described by the CK solution, either in the whole range of $\R$, as in 1D Fokker Planck approaches \citep[\textit{e.g.}][]{vasilievNewFokkerPlanckApproach2017, stoneRatesStellarTidal2016} or as the boundary condition at the loss cone interface for 2D models \citep[\textit{e.g.}][]{broggiExtremeMassRatio2022}. It is of primary interest to understand how these results should be interpreted when relaxing the hypothesis of instantaneous disruption.

The efficiency of two-body relaxation in altering the orbital parameters when $q \gtrsim 1$ implies that the classically computed rates are reliable in this regime, and should be computed with a loss cone radius slightly larger than $\eta_\mathrm{TTDE} \, r_\mathrm{TDE}$ accounting for the possibility of ejections due to mass loss ($r_\LC^A \simeq 0.8\, r_\mathrm{TDE}$ and $r_\LC^B \simeq 1.2\, r_\mathrm{TDE}$ for the models we considered). Among the events computed this way, one should identify the pericentres resulting into observable events, as discussed in \citet{bortolasPartialStellarTidal2023}.

The case of partial disruptions in the empty loss cone regime instead requires a more careful treatment. In the most extreme case we considered, $q=0.001$, stars with pericentres up to $r_\mathrm{p} \simeq 4.5 \, r_\mathrm{TDE}$ for model A and $r_\mathrm{p} \simeq 2.05$ for model B will be affected by the presence of the loss cone. The trend emerging from our analysis is that at smaller value of $q$, therefore higher specific binding energy, the critical value of the pericentre, up to which orbits result in pTDEs, will increase due to the larger number of orbits allowed. This will correspond to a larger rate of partial disruptions, possibly repeating.
Moreover, both models predict a significant change of the specific binding energy for particles as they approach the identified total disruption zone, with higher values of $q$ corresponding to stronger alterations. In the case of model A stars, part of the classically predicted rate will therefore result in \textit{failed} disruptions, since particles will be moved to higher $q$ areas as the sequence of weak partial disruptions takes place. On the other hand, model B stars might increase their specific binding energy completing the sequence of weak disruptions. In both cases, the inclusion of partial disruptions in Fokker-Planck solvers requires a proper modelling of the phenomenon accounting for modifications to the Cohn-Kulsrud profile, see Eq.~\ref{eq:CK-R0}.

For all models, the inclusion of repeated partial disruptions implies that a fraction of the classically expected TDEs will actually correspond to a sequence of weak disruptions, that may be too weak to be detected \citep{rossiProcessStellarTidal2021, Stone2020}, but can still contribute to the growth of the MBH. The large number of consecutive desruptions with weak mass loss may result in a sequence of bursts of gravitational waves, possibly increasing the gravitational wave background from TDEs \citep{toscaniGravitationalWaveBackground2020}.

It has been estimated that most of the TDEs  originates in the empty loss cone regime, with a fraction approximately given by  \citep[][Eq. 29]{stoneRatesStellarTidal2016}
\begin{equation}
    f_\mathrm{empty} = 1- 0.22\, \left(\frac{M_\bullet}{10^8\, M_\odot}\right)^{-0.307}
\end{equation}
corresponding to $70\%$ of the total expected cosmological rate. A qualitative result of our analysis is that part of these classically expected events will correspond to non-detectable pTDEs, therefore possibly reconciling the observed discrepancy between observed and expected TDE rates \citep{Stone2020, vanVelzen18}, while maintaining the possibility that stellar captures contribute to the growth of MBHs.

\subsection{Caveats}
In the case of model A, we determined the mass lost at the pericentre passage through Eq.~\ref{eq:mass_loss_ryu}. This equation has been derived for a MESA star with the core hydrogen mass fraction of 0.5 \citep{Ryu2020a} on a parabolic orbit. On the other hand, the effect of tidal excitation has been included through a prescription valid for $n=3.0$ polytropes. The MESA profile and the polytropic model correspond to different tidal excitation, with MESA model subject to stronger oscillations as shown in \citet{generozovOverabundanceBlackHole2018} (the corresponding function $T(\beta)$ is larger by a factor of a few).

On the other hand, when considering model B, we used the estimate for $\delta m$ from the simulations by \citet{guillochonHydrodynamicalSimulationsDetermine2013}. These do not account for relativistic effects, that are increasingly important as the mass of the central MBH grows and the pericentre of the orbit of partial disruption decreases.

\citet{liuRapidEvolutionRecurrence2024} simulated the partial disruption of different stars on orbits with eccentricity $e \simeq 0.95$ around an MBH with mass $M_\bullet=10^5 \, M_\odot$. The simulations include the disruption of a $1\, M_\odot$ star with radius $1 R_\odot$ initially described by a MESA model (corresponding to our model A stars). At small penetration factors, corresponding to weak mass loss, the simulated stars get slightly more tightly bound to the central MBH, the opposite of our predictions, showing a trend similar to what we expect for the less concentrated model B stars. In addition to the inconsistency between the MESA profile and the polytrope, this qualitative difference may also arise from the specific orbital parameters. In fact Eq.\ref{eq:mass_loss_ryu} (the function relating the pericentre distance and the mass lost at the pericentre passage)
has been fit to the simulation of disruptions on $e \gtrsim 0.99$ orbits \citep{Ryu2020a,Ryu2020b,Ryu2020c}, and the tidal excitation prescription similarly assume a parabolic orbit \citep{leeCrossSectionsTidal1986,ivanovNewModelTidally2001}. In general, the time spent at the pericentre increases as the eccentricity decreases, affecting both asymmetric mass loss and tidal excitation.

A general assumption in our treatment of repeating partial disruptions is that the stellar profile evolves self-similarly - with the same radius and the same polytropic index $n$. However, it is natural to expect that strong non-linear excitation and significant mass loss will result in a readjustment of the stellar profile and change, for example, the best fitting polytropic index. In general, when mass loss is extremely weak the outer layers of the star may expand to locally compensate the stripping. Moreover, as tidal oscillations become stronger, it is possible that part of the energy they store may thermalise and trigger a readjustment of the stellar profile, possibly by enlarging the radius of the star. In Appendix \ref{ap:therm} we explore this possibility and show that this effect mainly affects stars described by model B. In fact, being dominated by tidal excitation effects at larger $r_\mathrm{p}$, model B stars can end up in a total disruption at slightly larger pericentre compared to the reference scenario. Overall, the qualitative picture we discussed seems generally insensitive to stellar expansion, but we caution that our model for this is approximate and the matter should be investigated further with hydrodynamical simulations.

The model we built for partial disruptions holds only on a range of MBH masses. Repeated disruptions at small penetration factors may completely disrupt a star as long as the motion of the central MBHs around the centre of the system is negligible with respect to the orbital timescales, setting a lower bound to the mass of the central MBH. On the other hand, the star can undergo a disruption as long as it is not gravitationally captured before the disruption, setting an upper bound to the mass of the central black hole ($M_\bullet \lesssim 10^8 M_\odot$ for a sun-like star, the Hills mass). Finally, when the mass of the central black hole is large, one should account for general relativistic effects affecting the disruption process \citep{coughlinImpactRelativisticGravity2022} and the tidal excitation of the surviving core \citep{ivanovNewModelTidally2003}.

Finally, we included relaxation in our model only by setting the maximum number of pericentre passages that a repeating pTDE can perform. However, a more complete treatment should include discrete orbital fluctuations (in energy and angular momentum), \textit{e.g.} in a Monte Carlo fashion \citep{shapiroStarClustersContaining1978}. This will result, for example, in the possibility of interrupting the sequence of weak partial disruptions with a stronger event even in the empty loss cone regime \citep{weissbeinHowEmptyEmpty2017}.

\newpage
\subsection{Summary and Conclusions}

We have constructed a simple model for the partial disruption of a star around a massive black hole. We estimated of the number of consecutive partial disruptions that a star will undergo depending on the strength of two-body encounters, and estimated the final outcome of the sequence of disruptions as a function of the initial pericentre distance. We considered two stellar models of $m_\star = 1 \, M_\odot$ and $m_\star = 0.3\,M_\odot$, taking these two values as roughly representative of lower-main sequence and upper-main sequence internal structures.  We then built a model for partial disruptions that accounts for the effects of mass loss and tidal excitation of the stellar structure. We built our model based on the results of hydrodynamical simulations \citep{Ryu2020a,Ryu2020b,Ryu2020c, Ryu2020d, guillochonHydrodynamicalSimulationsDetermine2013, gaftonRelativisticEffectsTidal2015} and previous theoretical studies \citep{kremerHydrodynamicsCollisionsClose2022, stoneConsequencesStrongCompression2013, ivanovNewModelTidally2001}.
For each model we studied the problem of single partial disruptions, arguing that this is the relevant problem for loss cone orbits ($r_\mathrm{p}$ comparable to $r_\mathrm{TDE}$) with binding energy small enough that two-body relaxation is efficient, i.e. in the full loss cone regime.
We then considered the phenomenon of repeated partial disruptions, estimating the number of passages as the inverse of the loss cone diffusivity $q$, as in Eq.~\eqref{eq:npassages}. Therefore, pTDEs are expected to occur in the empty loss cone regime.

The main results of our work are summarised below:
\begin{itemize}
 \item The phenomenon of ejections, that is caused by significant mass loss in partial disruptions, may effectively enlarge the loss cone radius by 20\% compared to $\eta_\mathrm{TTDE}\, r_\mathrm{TDE}$.
 \item Partial disruptions may alter the orbital energy without destroying the stars. For stars with $m_\star=1\, M_\odot$, the energy change is driven by mass loss, and weak pTDEs happening at pericentres larger than $\sim 1.5 \,r_\mathrm{TDE}$ will result in a less tightly bound remnant. For stars with $m_\star=0.3\, M_\odot$, the energy change is driven by tidal oscillations, and weak pTDEs happening at pericentres larger than $\sim 1.6 \, r_\mathrm{TDE}$ will result in a more tightly bound remnant at the first pericentre passage.
 \item Stars in the empty loss cone are kicked by two-body relaxation towards orbits with smaller pericentres. Regardless of the stellar model, they are likely being consumed by a sequence of weak pTDEs. Only for $1\, M_\odot$ stars we find the possibility of being scattered onto a total TDE or away from the loss cone because of significant energy fluctuations when $q$ is small ($\lesssim 10^{-2}-10^{-3}$).
 \item TDEs originating in the empty loss cone may correspond to unobservable events. Since 70 \% of the TDEs are expected to originate in the empty loss cone regime, accounting for repeating pTDEs may reconcile expectations and observations of the total observed TDE rate.
\end{itemize}

We conclude by noting that our results strongly depend on the interplay of mass loss effects and tidal excitation. The simulation of a star being repeatedly disrupted by a MBH is currently too computationally expensive, since one needs to describe the hydrodynamical structure of the star including self gravity. The large computational cost is primarily due to the need of following a star along its nearly-radial orbit. However, a better understanding of repeating pTDEs offers a promising tentative solution for the tension between observational TDE rates and theoretical estimates.

\section*{Acknowledgements}
LB thanks David Izquierdo Villalba, Alessandro Lupi, and Martina Toscani for useful discussions.
We thank Julian Krolik for useful comments.
We thank the anonymous referee for their useful comments and suggestions.
MB acknowledges support provided by MUR under grant ``PNRR - Missione 4 Istruzione e Ricerca - Componente 2 Dalla Ricerca all'Impresa - Investimento 1.2 Finanziamento di progetti presentati da giovani ricercatori ID:SOE\_0163'' and by University of Milano-Bicocca under grant ``2022-NAZ-0482/B''.
LB, EB and AS acknowledge the financial support provided under the European
Union's H2020 ERC Consolidator Grant ``Binary Massive Black Hole Astrophysics'' (B Massive, Grant Agreement: 818691).
EB acknowledges support from the European Union's Horizon Europe programme under the Marie Skłodowska-Curie grant agreement No 101105915 (TESIFA), from the European
Consortium for Astroparticle Theory in the form of an Exchange Travel Grant, and the European Union’s Horizon 2020 Programme under the AHEAD2020 project
(grant agreement 871158).  NCS was supported by the Israel Science Foundation (Individual Research Grant 2565/19) and the Binational Science Foundation (grant Nos. 2019772 and 2020397).\\
Software: the Julia language \citep{bezansonJuliaFreshApproach2017}; Makie.jl \citep{danischMakieJlFlexible2021}; Roots.jl \citep{Roots.jl}; Interpolations.jl \citep{kittisopikulJuliaMathInterpolationsJl2023}.

\begin{appendix}
\begin{figure*}
 \includegraphics[width=0.49\linewidth]{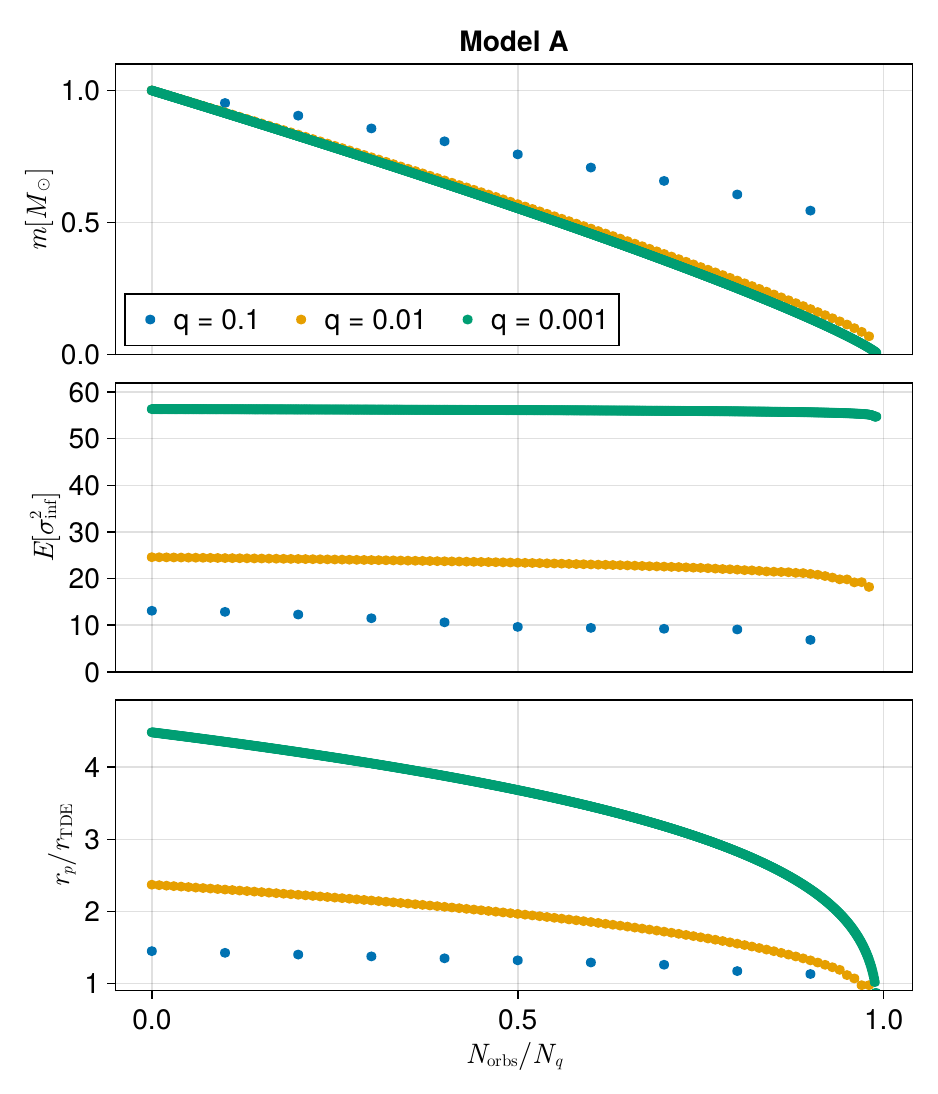}
 \includegraphics[width=0.49\linewidth]{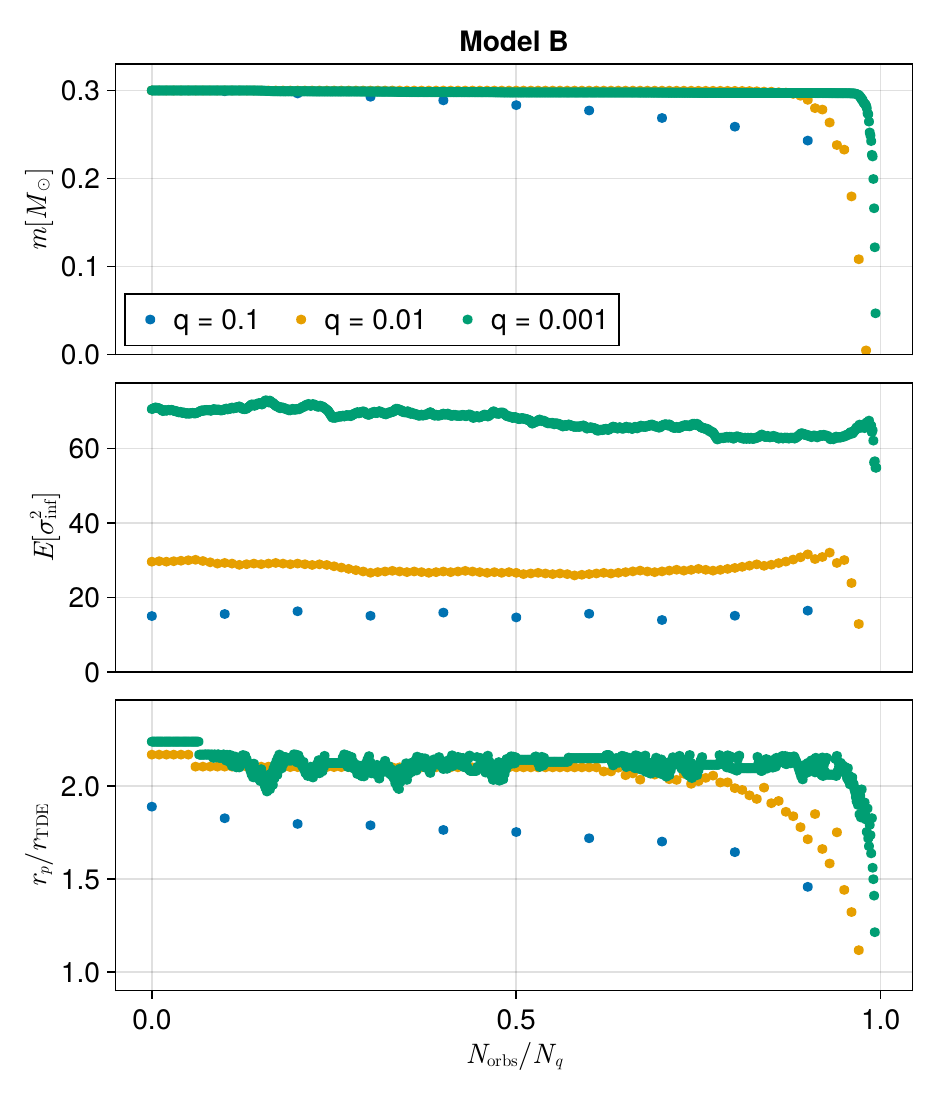}
\caption{\label{fig:app_threshold_orbits} Same as \ref{fig:threshold_orbits} but with the inclusion of the conversion of part of the oscillation energy $\epsilon_\mathrm{osc}$ into internal energy of the star through an enlargement of the stellar radius.}
\end{figure*}
\section{Expansion due to tidal oscillations}\label{ap:therm}
We report here a set of simulations that allow for the conversion of energy stored in oscillations into internal energy of the star. This is implemented through an expansion of the star, computed between step 3 and step 4 of the procedure outlined in \ref{subsec:repeated}. This is expected to happen only when the oscillations of the stellar structure are strongly non-linear. To estimate when this happens, we compare the total energy stored in oscillations with the internal energy of the star. For a polytrope model with index $n$ \citep{chandrasekharIntroductionStudyStellar1957}
\begin{equation}
   \epsilon_\mathrm{int} = -\frac{3}{5-n}\frac{G\, m_\star^2}{R_\star}\, .
\end{equation}
When $\epsilon_\mathrm{osc} > 3\% \, |\epsilon_\mathrm{int}|$ (an arbitrary threshold) we enlarge the stellar radius by a quantity
\begin{equation}
 \delta R_\star = R_\star \, \frac{\epsilon_\mathrm{osc}}{\epsilon_\mathrm{int}}
\end{equation}
and update $\epsilon_\mathrm{osc}$ consistently
\begin{equation}
 \delta \epsilon_\mathrm{osc} = - \delta \epsilon_\mathrm{int}.
\end{equation}
\begin{figure*}
    \centering
        \includegraphics[width=0.49\linewidth]{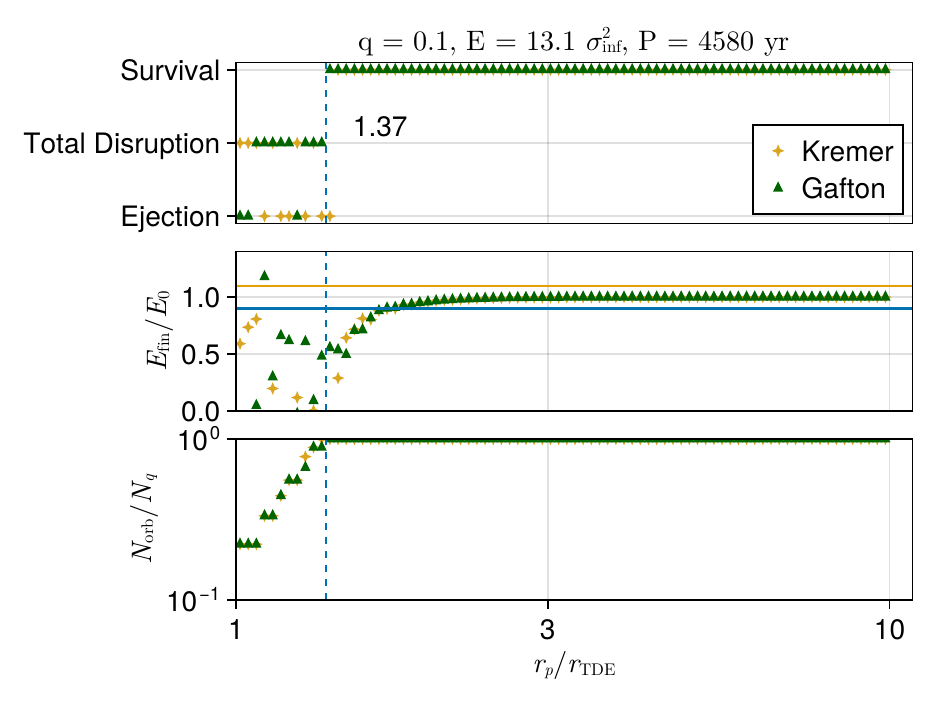}
        \includegraphics[width=0.49\linewidth]{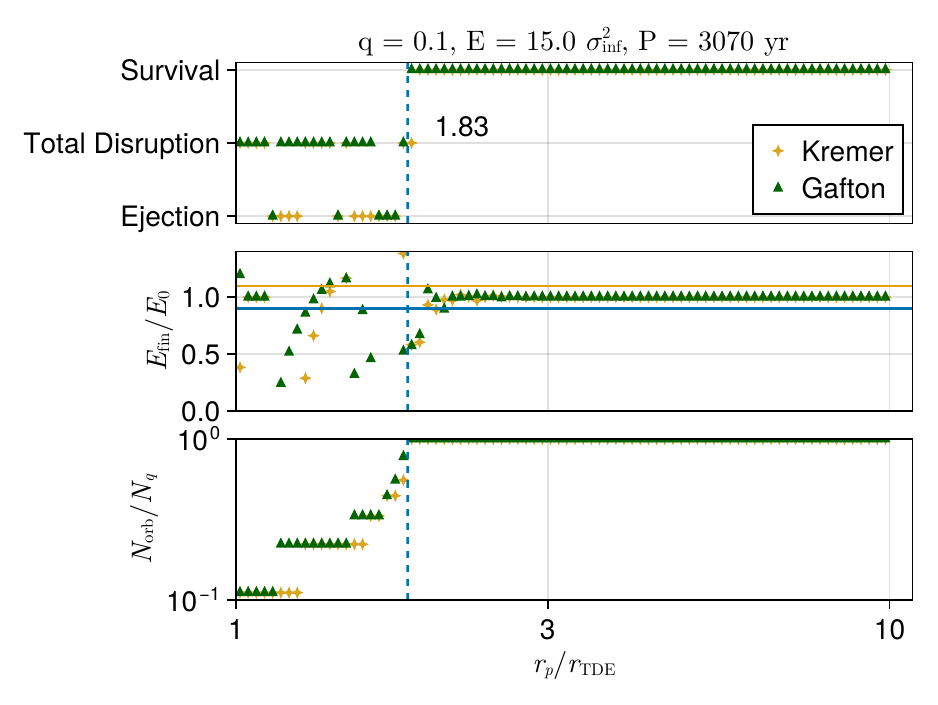}\\
        \includegraphics[width=0.49\linewidth]{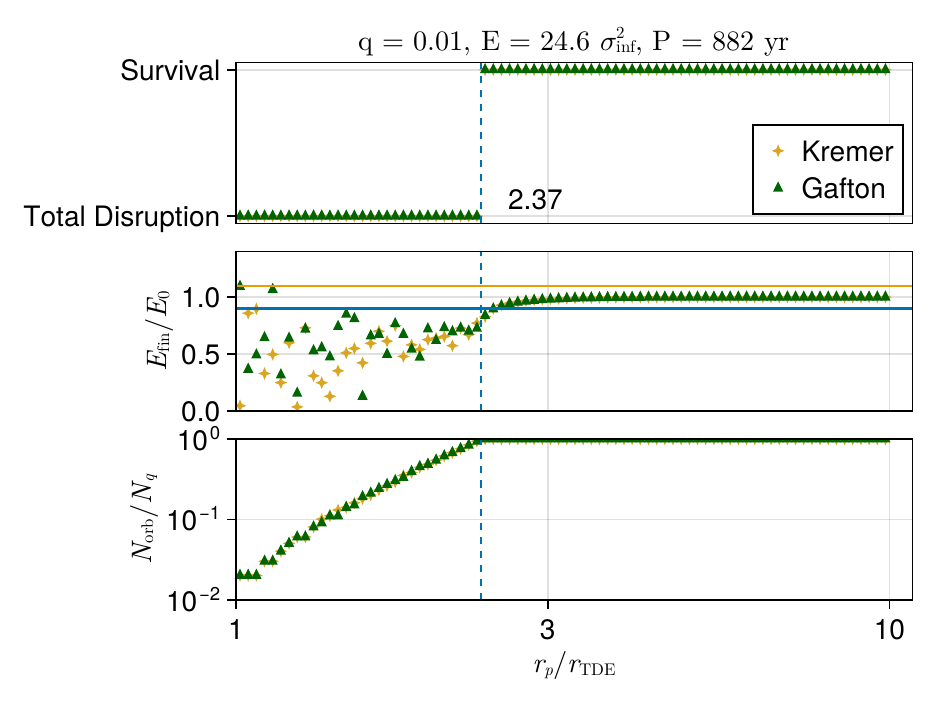}
        \includegraphics[width=0.49\linewidth]{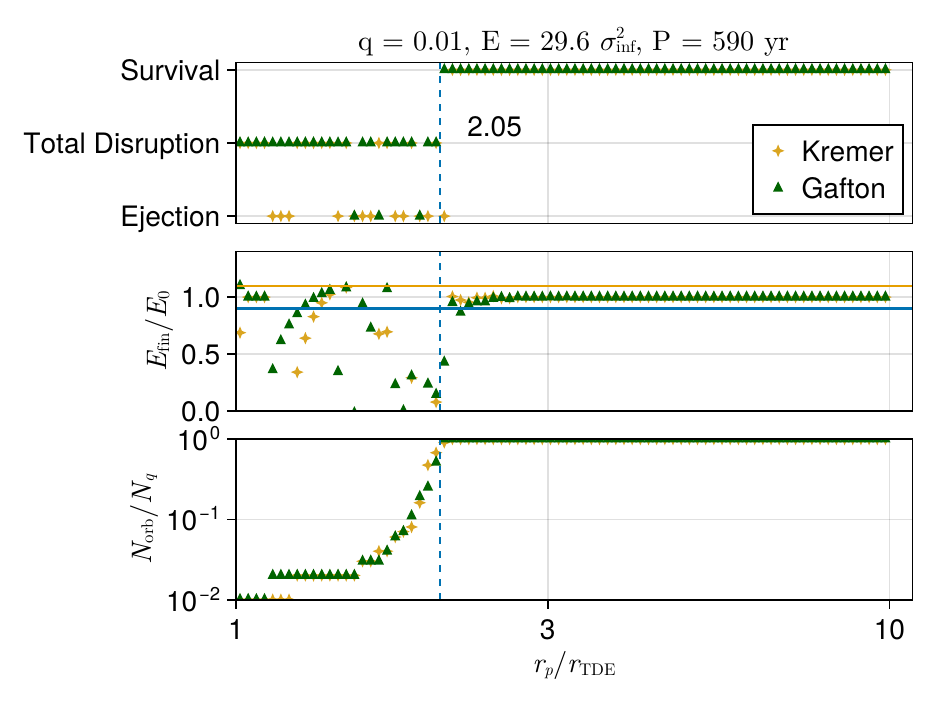}\\
        \includegraphics[width=0.49\linewidth]{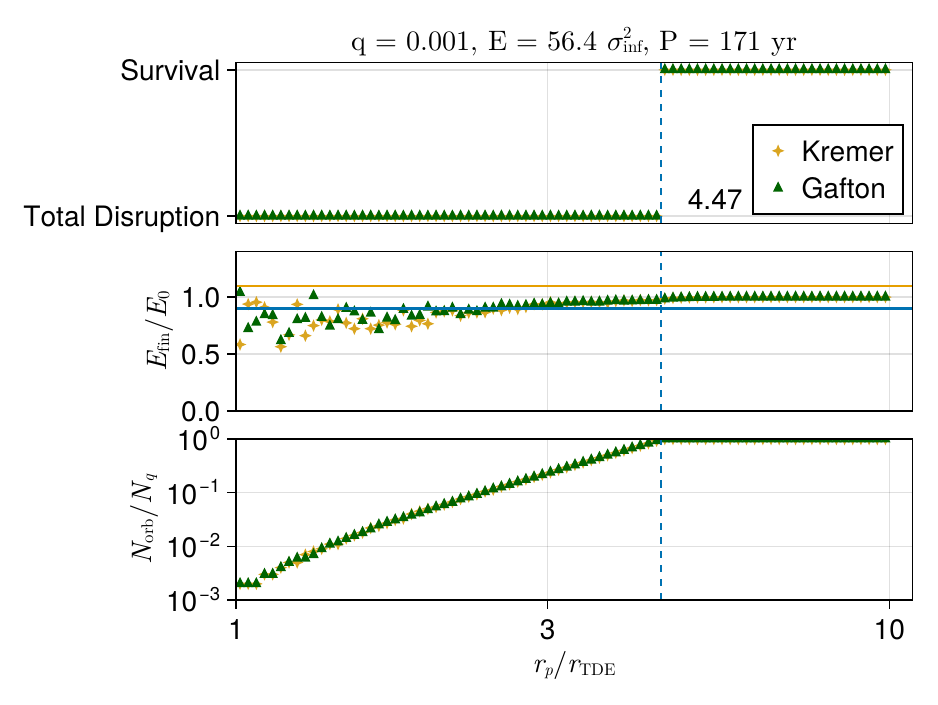}
        \includegraphics[width=0.49\linewidth]{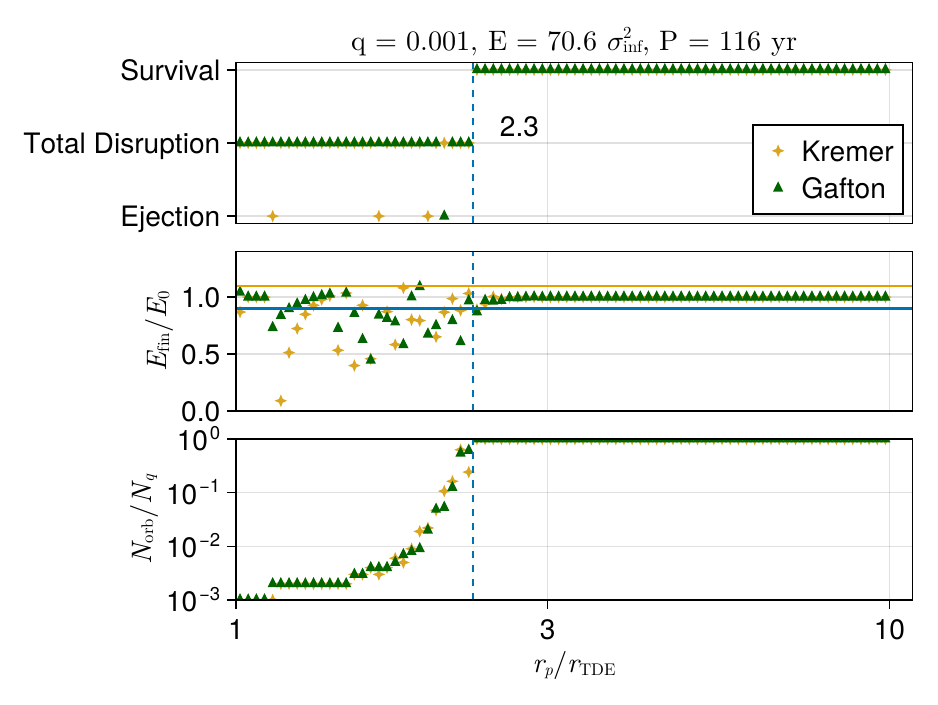}
        \caption{\label{fig:app_q_and_PTDE}Same as Fig.~\ref{fig:q_and_pTDE} but with the inclusion of the conversion of part of the oscillation energy $\epsilon_\mathrm{osc}$ into internal energy of the star through an enlargement of the stellar radius.}
\end{figure*}
In Fig.~\ref{fig:app_threshold_orbits} and \ref{fig:app_q_and_PTDE} we report the orbits at the threshold value of $r_\mathrm{p}$ and the overview of the outcomes at $q=0.1$, $q=0.01$ and $q=0.001$.

Compared to the reference model presented in the text, partial disruptions will completely destroy stars up to larger pericentre values. The effect is evident for model B stars, since they are strongly affected by tidal excitation at larger $r_\mathrm{p}$. Qualitatively, the results of this simulations are consistent with the conclusions presented in the main text.
\end{appendix}

	%
	\clearpage  
	\bibliographystyle{aa_url} 
	\bibliography{Biblio} 
	%

\end{document}